\documentclass[structabstract]{aa}

\usepackage{graphicx}
\usepackage[usenames]{color}

\newcommand{\vsini}{$v$\,sin\,$i$}

\begin{document}

\title{
Multiplicity of late-type B stars with HgMn peculiarity\thanks{Based on
observations obtained at the European Southern Observatory,
Paranal, Chile (ESO programme Nos.~074.D-0374 and 076.C-0170).}
}

   \author{M.~Sch\"oller\inst{1}
          \and
          S.~Correia\inst{2}
          \and
          S.~Hubrig\inst{2}
          \and
          N.~Ageorges\inst{3}
          }

   \institute{European Southern Observatory,
              Karl-Schwarzschild-Str.~2,
              85748 Garching, Germany\\
              \email{mschoell@eso.org}
         \and
             Astrophysikalisches Institut Potsdam,
             An der Sternwarte 16,
             14482 Potsdam, Germany
          \and
             Max-Planck-Institut f\"ur extraterrestrische Physik,
             Giessenbachstrasse 1, 
             85748 Garching, Germany
             }

   \date{Received September 15, 1996; accepted March 16, 1997}

 
  \abstract
   {
Observations at various wavelengths
of late B-type stars exhibiting strong overabundances
of the chemical elements Hg and Mn in their atmospheres
indicate that these stars are frequently
found in binary and multiple systems.
   }
   {
We intend to study the multiplicity of this type of chemically peculiar stars, 
looking for visual companions in the range of angular separation between
0\farcs05 and 8\arcsec{}.
   }
   {
We carried out a survey of 56 stars using diffraction-limited
near-infrared imaging with NAOS-CONICA at the VLT.
   }
   {
Thirty-three companion candidates in 24 binaries, three triples, and one quadruple system were detected.
Nine companion candidates were found for the first time in this study.
Five objects are likely chance projections.
The detected companion candidates have K magnitudes between 5\fm95 and 18\fm07
and angular separations ranging from $<$0\farcs05 to 7\farcs8,
corresponding to linear projected separations of 13.5--1700\,AU.
   }
   {
Our study clearly confirms that HgMn stars are frequently members of binary and multiple systems.
Taking into account companions found by other techniques,
the multiplicity fraction in our sample may be as high as 91\%.
The membership in binary and multiple systems seems to be
a key point to understanding the abundance patterns in these stars.
   }


   \maketitle

\section{Introduction}
\label{sect:intro}

Chemically peculiar (CP) stars are main-sequence A and B type stars
in the spectra 
of which lines of some elements are abnormally strong or weak. 
The class of CP stars is roughly represented by three subclasses:
the magnetic Ap and Bp stars, the metallic-line Am stars, and the HgMn stars,
which are late B-type stars
showing extreme overabundances of Hg (up to 6\,dex) and/or Mn (up to 3\,dex).

About 150 stars with HgMn peculiarity are currently known
(Renson \& Manfroid \cite{RensonManfroid2009}).
Most of them are rather young objects found
in young associations
such as Sco-Cen, Orion OB1, or Auriga OB1.
In contrast to classical Bp and Ap 
stars with large-scale organized magnetic fields, 
HgMn stars generally do not show overabundances of rare earth elements, but exhibit strong
overabundances of heavy elements such as W, Re, Os, Ir, Pt, Au, Hg, Tl, Pb, or Bi.
Another important distinctive feature of these stars is their slow rotation 
($\langle v\,\sin i\rangle \approx$ 29\,km\,s$^{-1}$,
Abt et~al.\ \cite{Abt1972a}).
The number of HgMn stars decreases sharply with increasing rotational
velocity (Wolff \& Wolff \cite{WolffWolff1974}).
Evidence that stellar rotation does affect 
abundance anomalies in HgMn stars is provided by the rather sharp cutoff in 
these anomalies at a 
projected rotational velocity of 70--80~km$\,$s$^{-1}$
(Hubrig \& Mathys \cite{HubrigMathys1996}).

The mechanisms responsible for
the development of the chemical anomalies of HgMn stars are not yet
fully understood. 
The abundance patterns may however be connected with binarity and multiplicity.
More than 2/3 of the HgMn stars are known
to belong to spectroscopic binaries
(Hubrig \& Mathys \cite{HubrigMathys1995}).

Over the past few years, we have been involved in 
extensive spectroscopic studies of upper main-sequence spectroscopic binaries with 
late B-type primaries  
with the goal of understanding why the vast majority of these stars exhibit in their 
atmospheres certain chemical abundance anomalies. 
Moreover, elements with anomalous abundances are  distributed inhomogeneously over the stellar surface
(e.g., Adelman et al.\ \cite{Adelman2002}; Hubrig et al.\ \cite{Hubrig2006}).
For $\alpha$\,And, secular evolution was found for Hg by Kochukhov et al.\ (\cite{Kochukhov2007}),
while a dynamical evolution of Ti, Sr, and Y spots within two months was reported by
Briquet et al.\ (\cite{Briquet2010}) for the spectroscopic binary HD\,11753.
Since a number of HgMn stars in binary systems is found at the zero age main-sequence (ZAMS;
e.g., Nordstrom \& Johansen \cite{NordstromJohansen1994}; Gonz\'alez et al.\ \cite{Gonzalez2006}), it is very likely that 
the timescale for developing a HgMn peculiarity is very short. 

Quite a number of HgMn stars belong to triple
or even quadruple systems
(Cole et~al.\ \cite{Cole1992}; Isobe \cite{Isobe1991}). 
Out of 30 SB HgMn stars observed with speckle interferometry,
15 appear to have more than two components.
Indirect evidence of a third component was found in
four other HgMn SBs (HD\,11905, HD\,34364, HD\,78316, and HD\,141556) on the
basis of spectroscopic and photometric arguments.
Additional evidence that other HgMn stars are frequently members of
multiple systems is inferred from the results of the ROSAT all-sky
survey.
X-ray emission was detected by this survey in twelve HgMn stars
(seven SB1s, three SB2s, and two for which no radial velocity data are available;
Bergh\"ofer et al.\ \cite{Berghoefer1996}, Hubrig \& Bergh\"ofer \cite{HubrigBerghoefer1998}).
Previous X-ray observations with the Einstein Observatory and
theoretical estimates had suggested that stars in the spectral range
B2--A7 are devoid of any significant X-ray emission. 
In most cases when emission had been detected in these stars, it was found to
originate in a cool companion.
This suggests that the X-ray
emission found in HgMn SBs does not originate in the HgMn primary.
From observations investigating late-B type stars with
X-ray emission using the ESO 3.6-m telescope with ADONIS,
Hubrig et~al.\ (\cite{Hubrig2001})
found faint companions to four HgMn stars that were part of the X-ray selected
late-B type stars observed, strengthening this interpretation.  

In the following, we report the results of our multiplicity study
of this class of objects using NACO K-band imaging.

\section{Observations and data reduction}
\label{sect:observations}

We carried out observations of 56 HgMn stars with NAOS-CONICA
(NACO; Lenzen et al.\ \cite{Lenzen2003}; Rousset et al.\ \cite{Rousset2003})
on the VLT in service
mode between October 2004 and March 2005, and again between November 2005 and
February 2006.
We used the S13 camera of NACO, which provides the smallest available pixel scale of 13.3\,milliarcsec
and a field-of-view of 13\farcs6.
All data were collected through a $Ks$ filter in image
autojitter mode, where the object is observed at typically
20 different image positions with random offsets between them.
Since all our sources are bright in $V$, we used the visible wavefront sensor of NAOS.

\begin{table*}
\centering
\caption{
Objects studied in the program.
}
\label{tab:objects}
\begin{tabular}{rccccr @{$\pm$} l}
\hline
\hline
\multicolumn{1}{c}{HD} &
\multicolumn{1}{c}{Other} &
\multicolumn{1}{c}{V} &
\multicolumn{1}{c}{K$^a$} &
\multicolumn{1}{c}{Spectral} &
\multicolumn{2}{c}{Parallax} \\
\multicolumn{1}{c}{number} &
\multicolumn{1}{c}{Identifier} &
\multicolumn{1}{c}{} &
\multicolumn{1}{c}{} &
\multicolumn{1}{c}{Type} &
\multicolumn{2}{c}{[mas]} \\
\hline
1909   & HR\,89 &	 6.56 & 6.66 &  B9IVmn & 5.13 & 0.83 \\
7374   & HR\,364 &	 5.96 & 6.21 &  B8III & 6.52 & 0.79 \\
11753  & HR\,558 &	 5.11 & 5.17 &  A3V & 10.55 & 0.69 \\
14228  & HR\,674 &	 3.55 &  &  B8IV-V & 21.06 & 0.61 \\
19400  & HR\,939 &	 5.50 &  &  B8III/IV & 6.21 & 0.50 \\
21933  & HR\,1079 &	 5.75 & 5.88 &  B9IV & 9.18 & 0.87 \\
23950  & HR\,1185 &	 6.07 & 5.98 &  B8III & 10.14 & 0.90 \\
27295  & HR\,1339 &	 5.49 & 5.61 &  B9IV & 12.20 & 0.76 \\
27376  & HR\,1347 &	 3.55 &  &  B9V & 18.27 & 0.55 \\
28217  & HR\,1402 &	 5.87 & 5.64 &  B8IV & 6.70 & 0.96 \\
29589  & HR\,1484 &	 5.45 & 5.66 &  B8IV & 9.46 & 0.78 \\
31373  & HR\,1576 &	 5.79 & 5.97 &  B9V & 7.71 & 0.87 \\
32964  & HR\,1657 &	 5.10 & 5.20 &  B9V+... & 11.65 & 0.73 \\
33647  & HR\,1690 &	 6.67 & 6.86 &  B9Vn & 2.03 & 1.04 \\
33904  & HR\,1702 &	 3.28 &  &  B9IV\&... & 17.69 & 0.71 \\
34364  & HR\,1728 &	 6.14 & 6.27 &  B9.5V & 8.20 & 0.78 \\
34880  & HR\,1759 &	 6.41 &  &  B8III & 4.80 & 1.32 \\
35548  & HR\,1800 &	 6.56 & 6.60 &  B9sp... & 4.42 & 0.80 \\
36881  & HR\,1883 &	 5.63 & 4.96 &  B9IIImnp... & 2.23 & 0.82 \\
37752  & HR\,1951 &	 6.58 & 6.71 &  B8p & 4.64 & 0.96 \\
38478  & HR\,1985 &	 5.99 & 6.13 &  B8IIImnp... & 3.71 & 0.91 \\
42657  & HR\,2202 &	 6.20 & 6.33 &  B9mnp... & 5.52 & 1.09 \\
49606  & HR\,2519 &	 5.87 & 6.15 &  B7III & 3.52 & 0.83 \\
51688  & HR\,2605 &	 6.39 & 6.64 &  B8III & 2.72 & 0.88 \\
53244  & HR\,2657 &	 4.10 & 4.37 &  B8II & 8.11 & 0.63 \\
53929  & HR\,2676 &	 6.09 & 6.37 &  B9.5III & 4.60 & 0.78 \\
59067  & HR\,2859 &	 5.87 & 3.75 &  A0$^b$ & 3.23 & 1.39 \\
63975  & HR\,3059 &	 5.13 & 5.37 &  B8II & 7.76 & 1.02 \\
65949  & CD-60 1951 &	 8.37 & 8.37 &  B8/B9 & 3.60 & 2.72 \\
65950  & CD-60 1952 &	 6.87 & 6.76 &  B8III & 2.91 & 0.57 \\
66259  & CD-60 1998 &	 8.33 &  &  B9.5V &  2.91 & 0.57$^c$ \\
66409  & CD-60 2009 &	 8.41 & 8.33 &  B8IV/V & 2.91 & 0.57$^c$ \\
68099  & HR\,3201 &	 6.08 & 6.35 &  B6III & 3.74 & 0.99 \\
68826  & AO\,Vel &	 9.42 & 9.11 &  B9III &  \multicolumn{2}{r}{$^d$}  \\
70235  & HR\,3273 &	 6.43 & 6.59 &  B8Ib/II & 4.00 & 0.58 \\
71066  & HR\,3302 &	 5.62 & 5.84 &  A0IVmn & 8.25 & 0.48 \\
71833  & HR\,3345 &	 6.67 & 6.78 &  B8II & 2.43 & 0.82 \\
72208  & HR\,3361 &	 6.83 & 6.83 &  B9p... & 4.96 & 1.04 \\
73340  & HR\,3413 &	 5.78 & 6.04 &  B8p... & 6.99 & 0.44 \\
75333  & HR\,3500 &	 5.31 &  &  B9mnp... & 7.45 & 0.74 \\
78316  & HR\,3623 &	 5.24 & 5.46 &  B8IIImnp & 6.74 & 0.91 \\
90264  & HR\,4089 &	 4.95 & 5.31 &  B8V & 7.59 & 0.50 \\
101189 & HR\,4487 &	 5.14 & 5.13 &  B9IV & 10.98 & 0.57 \\
110073 & HR\,4817 &	 4.63 & 4.79 &  B8II/III & 9.19 & 0.85 \\
120709$^e$ & HR\,5210 &	 4.56 &  &  B5III & 10.96 & 0.88 \\
124740 & CD-40 8541 &	 7.86 & 7.83 &  Ap... & 4.86 & 1.13 \\
129174 & HR\,5475 &	 4.91 &  &  B9p+... & 10.28 & 0.91 \\
141556 & HR\,5883 &	 3.96 &  &  B9IV & 15.86 & 0.84 \\
144661 & HR\,5998 &	 6.32 & 6.43 &  B8IV/V & 8.50 & 0.84 \\
144844 & HR\,6003 &	 5.86 & 5.71 &  B9V & 7.65 & 0.77 \\
158704 & HR\,6520 &	 6.06 & 6.15 &  B9II/III & 7.47 & 1.10 \\
165493 & HR\,6759 &	 6.15 &  &  B7.5II & 4.00 & 1.01 \\
178065 & HR\,7245 &	 6.56 & 6.35 &  B9III & 4.34 & 0.88 \\
216494 & HR\,8704 &	 5.78 & 5.93 &  B8IV/V & 4.96 & 0.84 \\
221507 & HR\,8937 &	 4.37 & 4.61 &  B9.5IVmnpe... & 18.28 & 0.80 \\
224926 & HR\,9087 &	 5.12 & 5.44 &  B7III-IV & 7.98 & 0.71 \\
\hline
41040$^f$  & HR\,2130 &	 5.14 & 5.36 &  B8III & 3.05 & 0.96 \\
\hline
\end{tabular}  
\begin{flushleft}
Remarks:\\
$^a$Please note that for some targets, SIMBAD is not listing a K magnitude.\\
$^b$SIMBAD is listing a spectral type G8Ib-II+... for HD\,59067. We adopted the spectral type given by Schneider (\cite{Schneider1981}).\\
$^c$HD\,65949, HD\,65950, HD\,66259, and HD\,66409 are all members of the open cluster NGC\,2516.
Since SIMBAD does not provide an individual parallax,
we adopted the parallax of the brighter star HD\,65950 for HD\,66259 and HD\,66409.\\
$^d$Gonz\'alez et al.\ (\cite{Gonzalez2006}) give a distance of 0.72\,kpc for  HD\,68826 (AO\,Vel).\\
$^e$HD\,120709 belongs to the subgroup of PGa stars, which are usually considered as the hotter extension
of HgMn stars, exhibiting deficient He and strongly overabundant P and Ga (e.g.\ Castelli et al.\ \cite{Castelli1997}).\\
$^f$HD\,41040 is not known to present spectral peculiarities, but the orbital parameters of the short
period subsystem are in the range of those typical for HgMn stars.
It is not included in the sample of 56 HgMn stars for statistical considerations.
\end{flushleft}
\end{table*}

The observed sample was mainly selected from the ``Catalogue and Bibliography of Mn-Hg Stars''
(Schneider \cite{Schneider1981}) taking into account accessibility from the VLT.
A few targets not listed in this catalogue were selected on the basis of
previous spectroscopic studies of late B-type stars, in which HgMn peculiarity was detected.
The sample is listed in Table~\ref{tab:objects}.
In Col.~1, we give the HD number of the objects,
in Col.~2 another identifier, in Cols.~3 and 4 the magnitudes
in $V$ and $K$ bands,
in Col.~5 the spectral type,
and finally in Col.~6 the parallax.
All information was collected from the SIMBAD database.

The data reduction was performed with the eclipse package in the standard way. 
Sky background frames obtained from median averaging
of the jittered frames were subtracted from the individual frames.
All frames were then flat-fielded and corrected for bad pixels
using calibration files provided by ESO.

Astrometry and relative photometry of the 
multiple systems were performed using the IRAF package DAOPHOT.
We assume that systematics introduced are 1/10$^{\rm th}$ of a pixel for the position of
the individual objects and 1\% for the flux ratio,
which we used unless the errors determined by IRAF were higher, generally for
the faintest companion candidates.
The final errors in the relative positions are estimated by combining quadratically the 
rms variations in our astrometric analysis with the 
uncertainty in the plate scale (13.26$\pm$0.03\,mas)
and detector orientation ($\pm$0.5$^{\circ}$), both provided by Masciadri et al.\ (\cite{Masciadri2003}).

\section{Results}
\label{sect:results}

\begin{table*}
\centering
\caption{
Astrometric and photometric results of the candidate binaries and multiples resolved in our study.
}
\label{tab:astrometry}
{\scriptsize
\begin{tabular}{rrr @{$\pm$} rr @{$\pm$} rr @{$\pm$} rr @{$\pm$} rr @{$\pm$} rr @{$\pm$} rr @{$\pm$} rc}
\hline
\hline
\multicolumn{1}{c}{HD} &
\multicolumn{1}{c}{MJD} &
\multicolumn{2}{c}{Separation} &
\multicolumn{2}{c}{Position} &
\multicolumn{2}{c}{K mag } &
\multicolumn{2}{c}{K mag } &
\multicolumn{2}{c}{K mag } &
\multicolumn{2}{c}{K mag } &
\multicolumn{2}{c}{Projected} &
\multicolumn{1}{c}{Chance} \\
\multicolumn{1}{c}{number} &
\multicolumn{1}{c}{} &
\multicolumn{2}{c}{} &
\multicolumn{2}{c}{angle} &
\multicolumn{2}{c}{difference} &
\multicolumn{2}{c}{system} &
\multicolumn{2}{c}{primary} &
\multicolumn{2}{c}{secondary} &
\multicolumn{2}{c}{linear} &
\multicolumn{1}{c}{projection } \\
\multicolumn{1}{c}{} &
\multicolumn{1}{c}{} &
\multicolumn{2}{c}{} &
\multicolumn{2}{c}{} &
\multicolumn{2}{c}{} &
\multicolumn{2}{c}{} &
\multicolumn{2}{c}{} &
\multicolumn{2}{c}{} &
\multicolumn{2}{c}{separation} &
\multicolumn{1}{c}{probability} \\
\multicolumn{1}{c}{} &
\multicolumn{1}{c}{} &
\multicolumn{2}{c}{[\arcsec{}]} &
\multicolumn{2}{c}{[$^{\circ}$]} &
\multicolumn{2}{c}{} &
\multicolumn{2}{c}{} &
\multicolumn{2}{c}{} &
\multicolumn{2}{c}{} &
\multicolumn{2}{c}{[AU]} &
\multicolumn{1}{c}{[\%]} \\
\hline
\multicolumn{17}{c}{Binaries}\\
\hline
         21933 & 53376.06 & 0.124 & 0.003 & 112.2 & 1.7 & 2.43 & 0.01 & 5.88 & 0.02 & 5.99 & 0.02 & 8.42 & 0.04   & 13.5   & 1.3   & 4.24$\times10^{-6}$ \\
         21933 & 53700.14 & 0.149 & 0.003 & 111.9 & 1.4 & 2.70 & 0.01 & 5.88 & 0.02 & 5.97 & 0.02 & 8.66 & 0.04   & 16.2   & 1.6   & 6.19$\times10^{-6}$ \\
$^{\ast}$27376 & 53690.21 & 5.384 & 0.003 & 162.2 & 0.5 & 6.21 & 0.03 & 3.95 & 0.24 & 3.96 & 0.24 & 10.17 & 0.27  & 294.7  & 8.9   & 4.83$\times10^{-2}$ \\
$^{\ast}$28217 & 53409.04 & 0.119 & 0.004 & 27.3 & 1.8 & 1.24 & 0.01 & 5.64 & 0.02 & 5.94 & 0.03 & 7.19 & 0.05    & 17.8   & 2.6   & 3.96$\times10^{-6}$ \\
$^{\ast}$32964 & 53286.35 & 1.599 & 0.004 & 308.3 & 0.5 & 3.82 & 0.01 & 5.20 & 0.02 & 5.23 & 0.02 & 9.05 & 0.05   & 137.3  & 8.6   & 2.84$\times10^{-3}$ \\
$^{\ast}$33647 & 53347.21 & 0.147 & 0.003 & 345.1 & 1.4 & 0.79 & 0.01 & 6.86 & 0.02 & 7.29 & 0.03 & 8.07 & 0.05   & 72.4   & 37.1  & 1.80$\times10^{-5}$ \\
         33904 & 53374.17 & 0.352 & 0.003 & 250.9 & 0.7 & 3.29 & 0.01 & 3.52 & 0.24 & 3.58 & 0.24 & 6.86 & 0.26   & 19.9   & 0.8   & 3.43$\times10^{-5}$ \\
$^{\ast}$35548 & 53380.15 & 0.300 & 0.003 & 174.1 & 0.7 & 0.76 & 0.01 & 6.60 & 0.02 & 7.04 & 0.03 & 7.80 & 0.05   & 67.9   & 12.3  & 2.50$\times10^{-5}$ \\
$^{\ast}$36881 & 53348.22 & 2.801 & 0.003 & 351.9 & 0.5 & 4.16 & 0.01 & 4.96 & 0.02 & 4.98 & 0.02 & 9.14 & 0.05   & 1256.1 & 461.9 & 1.74$\times10^{-2}$ \\
$^{\ast}$42657 & 53741.18 & 0.688 & 0.003 & 202.9 & 0.6 & 1.45 & 0.01 & 6.33 & 0.03 & 6.58 & 0.03 & 8.03 & 0.06   & 124.6  & 24.6  & 5.26$\times10^{-4}$ \\
         53244 & 53408.14 & 0.332 & 0.004 & 114.8 & 0.8 & 4.26 & 0.01 & 4.37 & 0.04 & 4.39 & 0.04 & 8.65 & 0.06   & 40.9   & 3.2   & 1.84$\times10^{-4}$ \\
         53929 & 53408.17 & 3.659 & 0.003 & 345.5 & 0.5 & 3.76 & 0.01 & 6.37 & 0.02 & 6.40 & 0.02 & 10.16 & 0.04  & 795.4  & 134.9 & 7.44$\times10^{-2}$ \\
         $^{\ast}$59067 & 53379.26 & 0.811 & 0.003 & 170.4 & 0.5 & 3.56 & 0.01 & 3.75 & 0.04 & 3.79 & 0.04 & 7.34 & 0.06   & 251.1  & 108.1 & 3.65$\times10^{-4}$ \\
         72208 & 53404.25 & 0.671 & 0.004 & 332.1 & 0.6 & 6.24 & 0.02 & 6.83 & 0.03 & 6.84 & 0.03 & 13.08 & 0.06  & 135.3  & 28.4  & 6.12$\times10^{-3}$ \\
$^{\ast}$73340 & 53404.28 & 0.566 & 0.004 & 219.7 & 0.6 & 2.38 & 0.01 & 6.04 & 0.02 & 6.16 & 0.02 & 8.54 & 0.05   & 81.0   & 5.1   & 8.90$\times10^{-4}$ \\
$^{\ast}$75333 & 53379.30 & 1.316 & 0.003 & 167.5 & 0.5 & 3.79 & 0.01 & 5.44 & 0.02 & 5.47 & 0.02 & 9.26 & 0.05   & 176.6  & 17.6  & 4.81$\times10^{-4}$ \\
$^{\ast}$78316 & 53380.29 & 0.269 & 0.003 & 109.7 & 0.9 & 2.55 & 0.01 & 5.46 & 0.02 & 5.56 & 0.03 & 8.11 & 0.05   & 39.9   & 5.4   & 6.01$\times10^{-5}$ \\
         90264 & 53779.20 & 2.219 & 0.003 & 350.2 & 0.5 & 5.51 & 0.01 & 5.31 & 0.02 & 5.32 & 0.02 & 10.83 & 0.05  & 292.4  & 19.3  & 6.70$\times10^{-2}$ \\
         101189 & 53403.37 & 0.337 & 0.003 & 104.1 & 0.7 & 1.78 & 0.01 & 5.13 & 0.02 & 5.32 & 0.03 & 7.10 & 0.05  & 30.7   & 1.6   & 2.84$\times10^{-4}$ \\
$^{\ast}$110073 & 53404.31 & 1.202 & 0.003 & 73.9 & 0.5 & 3.12 & 0.01 & 4.78 & 0.02 & 4.84 & 0.02 & 7.96 & 0.04   & 130.8  & 12.1  & 8.03$\times10^{-4}$ \\
$^{\ast}$120709 & 53404.34 & 7.830 & 0.003 & 105.5 & 0.5 & 1.55 & 0.01 & 4.97 & 0.03 & 4.97 & 0.03 & 6.52 & 0.04  & 714.4  & 57.4  & 3.41$\times10^{-2}$ \\
$^{\ast}$129174 & 53404.40 & 5.537 & 0.003 & 110.5 & 0.5 & 0.27 & 0.01 & 5.05 & 0.02 & 5.67 & 0.03 & 5.95 & 0.05  & 538.6  & 47.7  & 8.52$\times10^{-3}$ \\
$^{\ast}$165493 & 53442.37 & 4.041 & 0.003 & 257.4 & 0.5 & 2.96 & 0.01 & 6.33 & 0.02 & 6.40 & 0.02 & 9.36 & 0.05  & 1010.3 & 255.1 & 8.62$\times10^{-2}$ \\
$^{\ast}$216494 & 53298.09 & 0.069 & 0.003 & 285.9 & 2.8 & 0.35 & 0.01 & 5.93 & 0.02 & 6.52 & 0.03 & 6.87 & 0.06  & 13.9   & 2.4   & 1.31$\times10^{-6}$ \\
         221507 & 53296.63 & 0.641 & 0.004 & 240.2 & 0.6 & 2.75 & 0.01 & 4.61 & 0.03 & 4.69 & 0.03 & 7.44 & 0.05  & 35.1   & 1.6   & 1.14$\times10^{-4}$ \\
\hline
$^{\ast}$41040 & 53348.25 & \multicolumn{2}{c}{$<$0.05} & 233.6 & 10.0 & \multicolumn{2}{c}{$\sim$0.01} & 5.36 & 0.02 & \multicolumn{2}{c}{} & \multicolumn{2}{c}{} & \multicolumn{2}{c}{$<$16} & 5.71$\times10^{-9}$ \\
\hline
\multicolumn{17}{c}{Triple systems}\\
\hline
$^{\ast}$34880AB & 53379.17 & 0.511 & 0.003 & 199.6 & 0.6 & 2.25 & 0.01 & 6.43 & 0.02 & 6.56 & 0.02 & 8.82 & 0.05            & 106.5  & 29.3  & 3.62$\times10^{-4}$ \\
$^{\ast}$34880AC & & 4.571 & 0.007 & 284.9 & 0.5 & 2.92 & 0.01 & \multicolumn{2}{c}{} & \multicolumn{2}{c}{} & 9.48 & 0.03   & 952.3  & 261.9 & 4.64$\times10^{-2}$ \\
$^{\ast}$158704AB & 53442.34 & 0.434 & 0.003 & 192.9 & 0.7 & 1.16 & 0.01 & 6.15 & 0.02 & 6.47 & 0.02 & 7.64 & 0.03           & 58.1   & 8.6   & 1.26$\times10^{-3}$ \\
         158704AC & & 1.637 & 0.004 & 123.6 & 0.6 & 6.18 & 0.03 & \multicolumn{2}{c}{} & \multicolumn{2}{c}{} & 12.65 & 0.03 & 219.1  & 32.3  & 2.33 \\
         178065AB & 53286.01 & 6.427 & 0.004 & 204.1 & 0.5 & 6.90 & 0.02 & 6.35 & 0.02 & 6.35 & 0.02 & 13.25 & 0.03          & 1480.9 & 300.3 & 16.7 \\
         178065AC & & 3.319 & 0.004 & 64.9 & 0.5 & 8.96 & 0.07 & \multicolumn{2}{c}{} & \multicolumn{2}{c}{} & 15.31 & 0.05  & 764.7  & 155.1 & $>$16.7 \\
\hline
\multicolumn{17}{c}{Quadruple system}\\
\hline
         66259AB & 53731.35 & \multicolumn{2}{c}{$<$0.05} & 2.5 & 10.0 & \multicolumn{2}{c}{$\sim$0.04} & 8.25 & 0.09 & \multicolumn{2}{c}{} & \multicolumn{2}{c}{} & \multicolumn{2}{c}{$<$17}   & 7.00$\times10^{-7}$ \\
         66259AC & & 4.861 & 0.006 & 110.8 & 0.5 & 8.18 & 0.08 & \multicolumn{2}{c}{} & \multicolumn{2}{c}{} & 17.16 & 0.11 & 1670.4 & 327.2 & $>$5.60 \\
         66259CD & & 0.415 & 0.013 & 65.9 & 1.8 & 0.90 & 0.02 & \multicolumn{2}{c}{} & \multicolumn{2}{c}{} & 18.07 & 0.12  & 142.6  & 28.0  & $>$6.29 \\
\hline
\end{tabular}  
}
\end{table*} 

We found in total 24 binaries, three triple systems, and one quadruple system in our survey.
Out of the 33 companion candidates, we infer that five are chance projections.
Nine companion candidates were detected for the first time.
The astrometric and photometric results are presented in Table~\ref{tab:astrometry} for all 
multiple systems of our sample. 
In Col.~1, we list the HD number for each system as well as the pair designation for
the higher order systems, and
Col.~2 gives the modified Julian date for the observations.
In Cols.~3, 4, and 5, we show the separation, position angle, and magnitude difference
in the K band between the components, as retrieved by aperture photometry from our images.
In Col.~6, we give the K band magnitude for the whole system, as derived from the 2MASS or
DENIS catalogues,
and in Cols.~7 and 8 we give K band magnitudes for the primary and secondary component, as
determined from Cols.~5 and 6.
Column~9 lists finally the chance projection probability of the secondary component,
as described in Sect.~\ref{sect:projections}.
An asterisk preceding the HD number in Col.~1 indicates systems where the companion was
known before our study.
For the two companion candidates that we find within the point spread function, we provide only rough
astrometric and photometric estimates.

\begin{figure*}
\centering
\includegraphics[width=0.24\textwidth, angle=0]{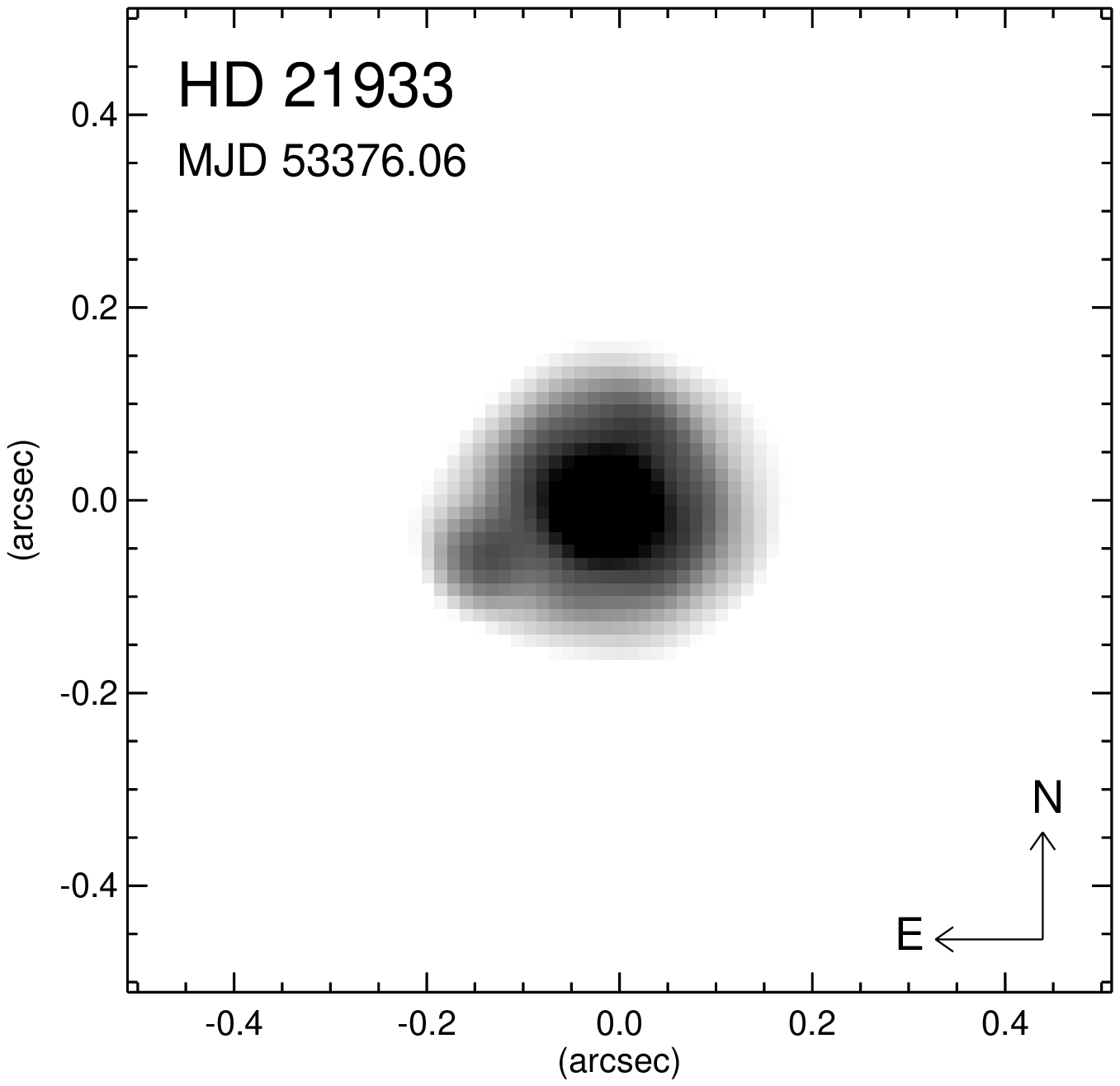}
\includegraphics[width=0.24\textwidth, angle=0]{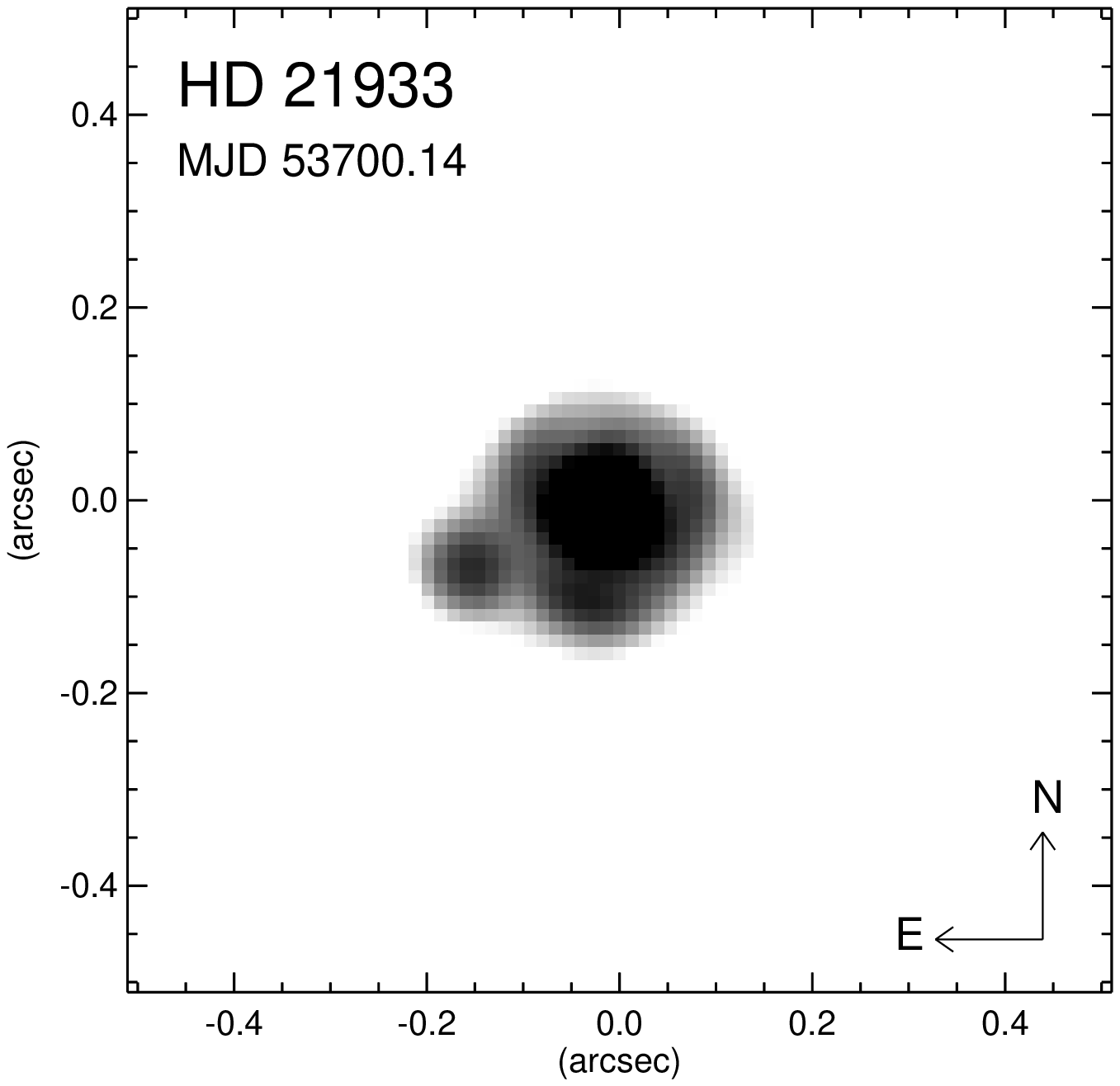}
\includegraphics[width=0.24\textwidth, angle=0]{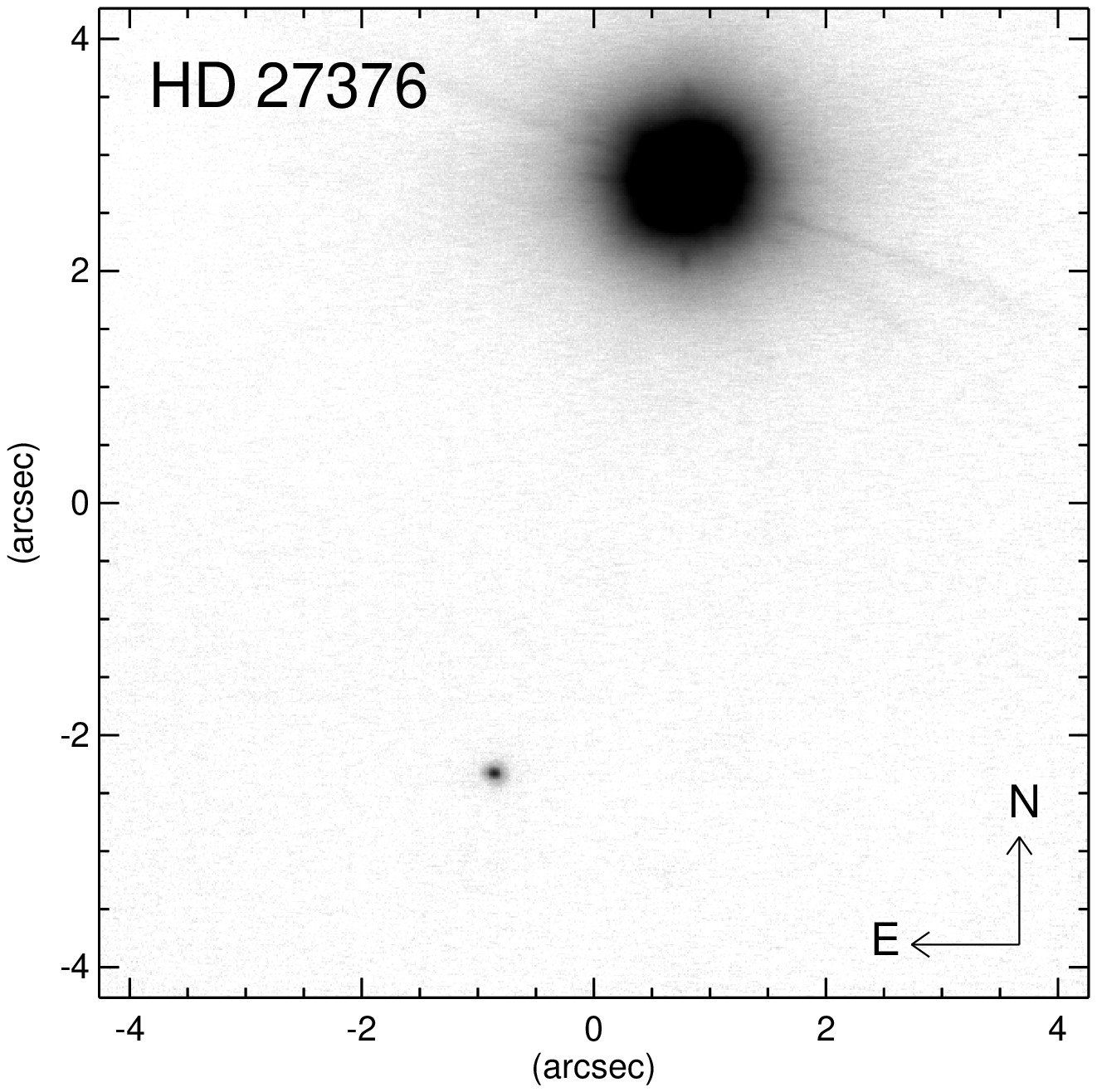}
\includegraphics[width=0.24\textwidth, angle=0]{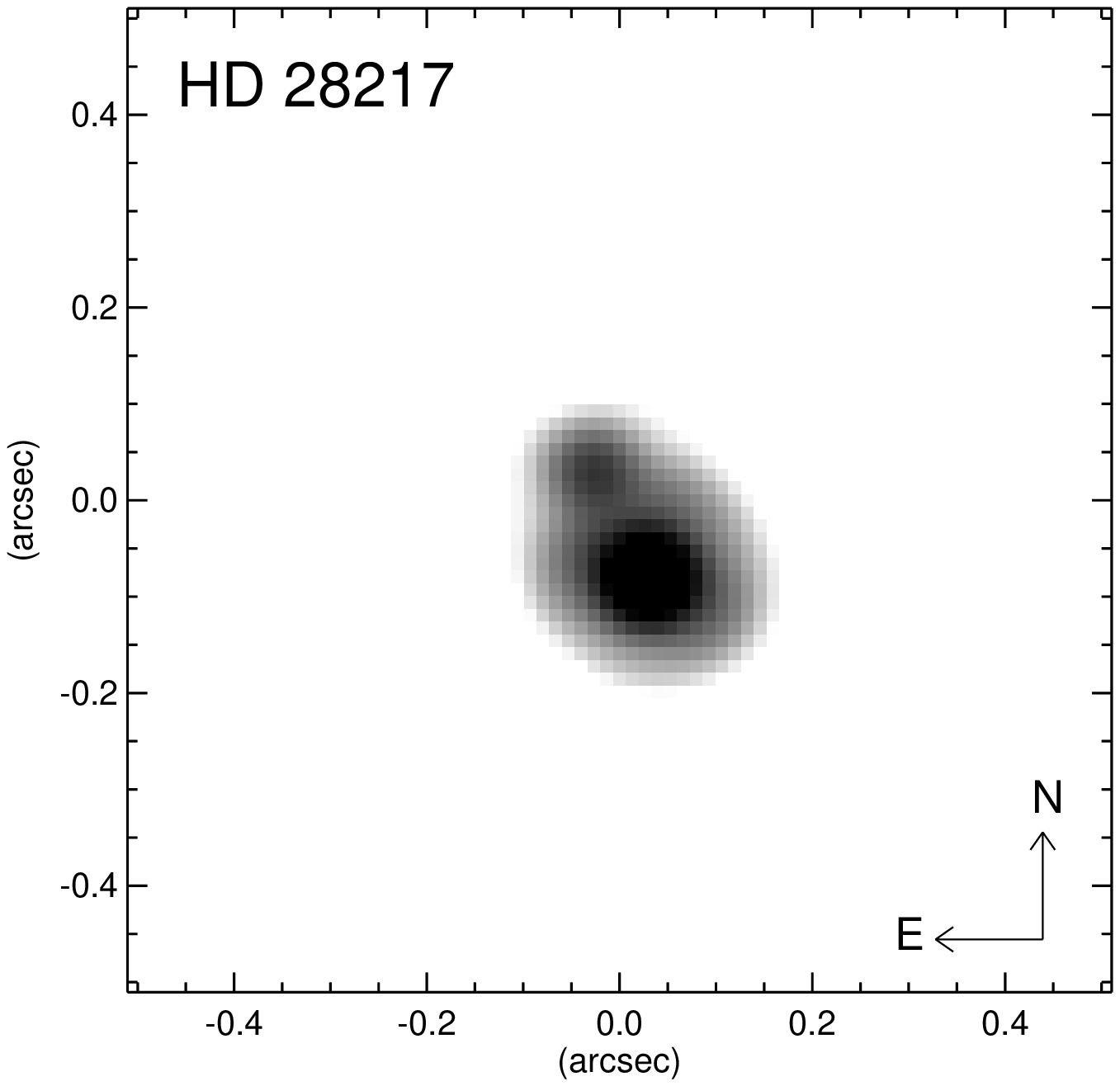}
\includegraphics[width=0.24\textwidth, angle=0]{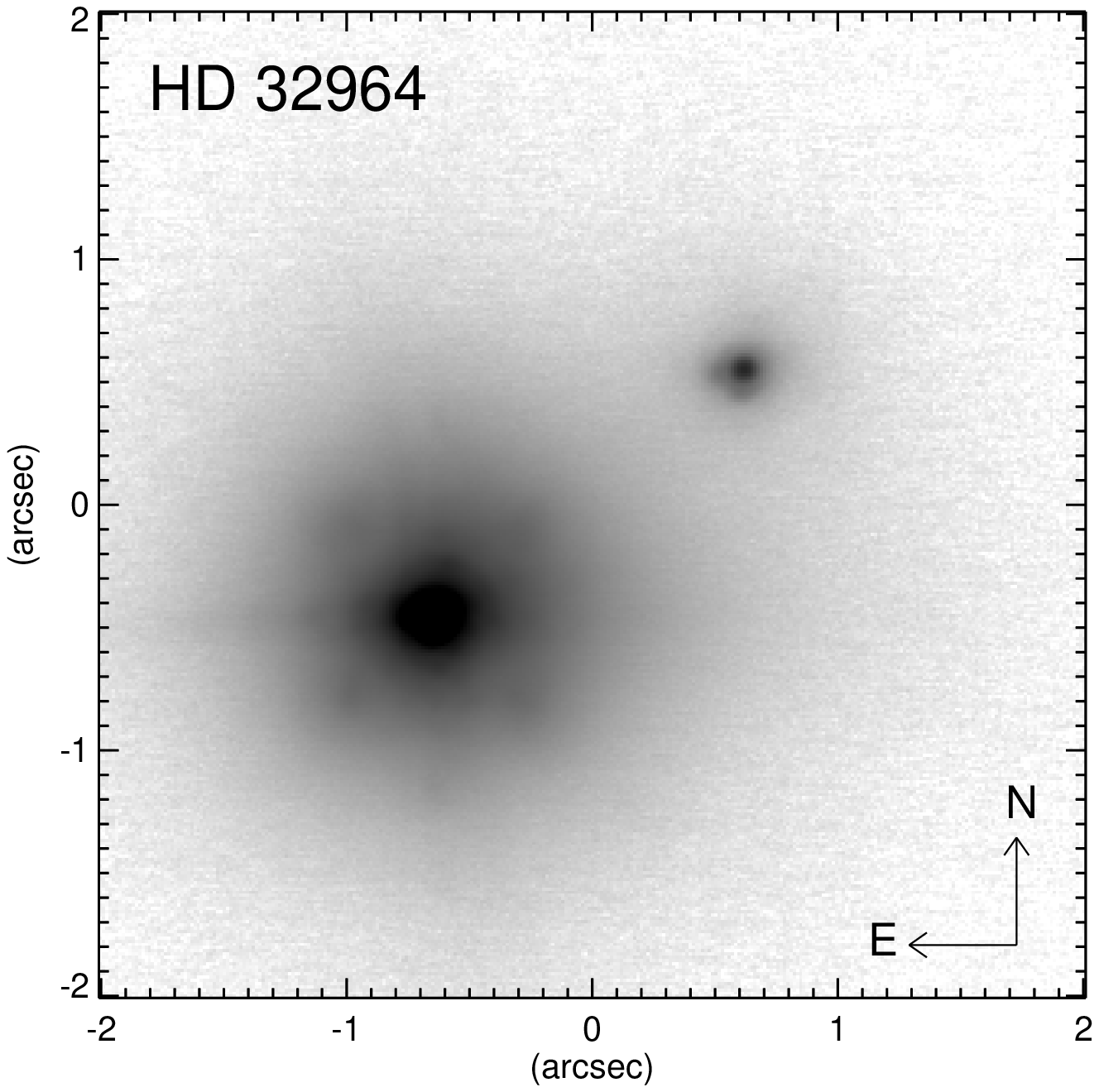}
\includegraphics[width=0.24\textwidth, angle=0]{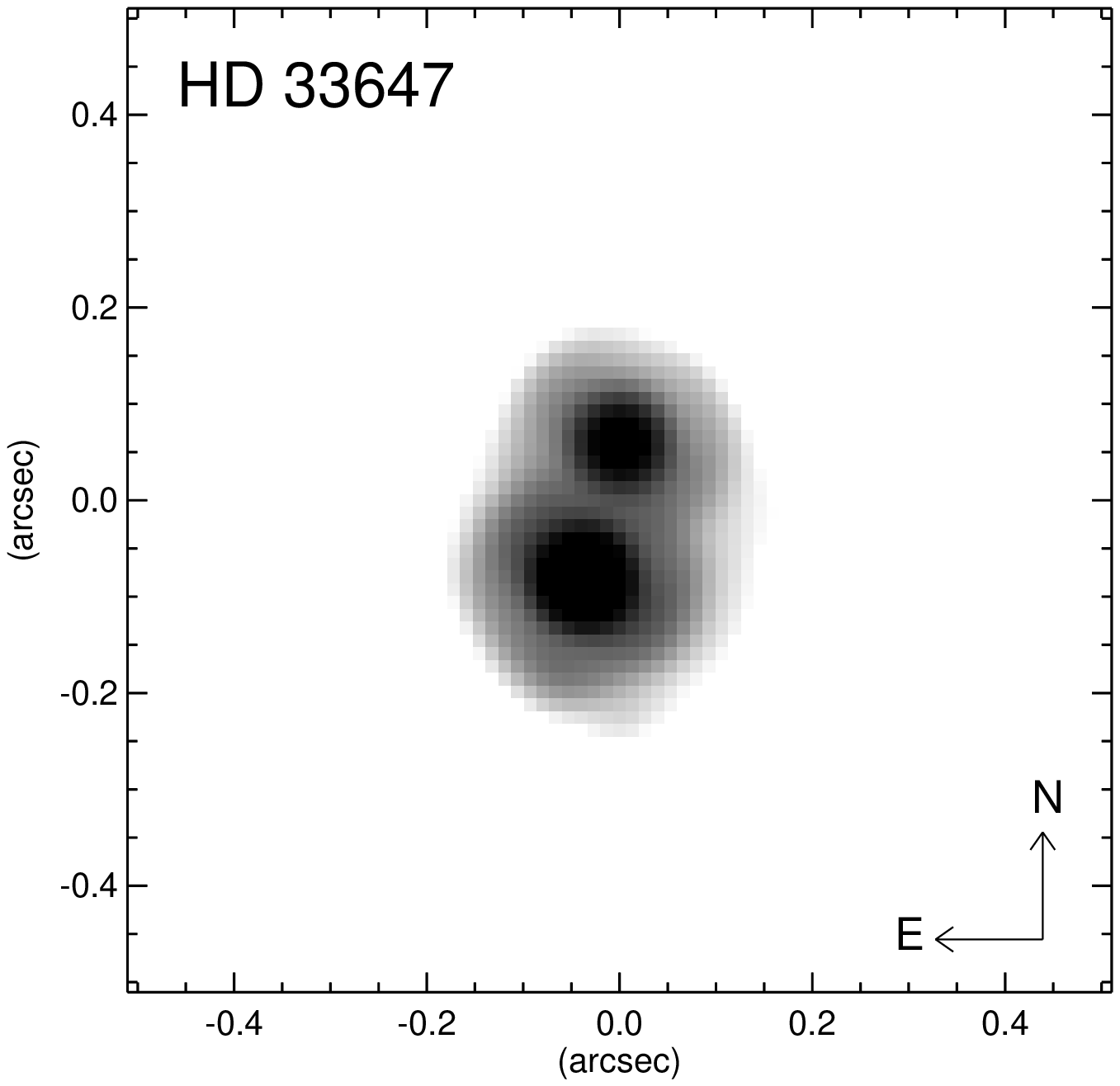}
\includegraphics[width=0.24\textwidth, angle=0]{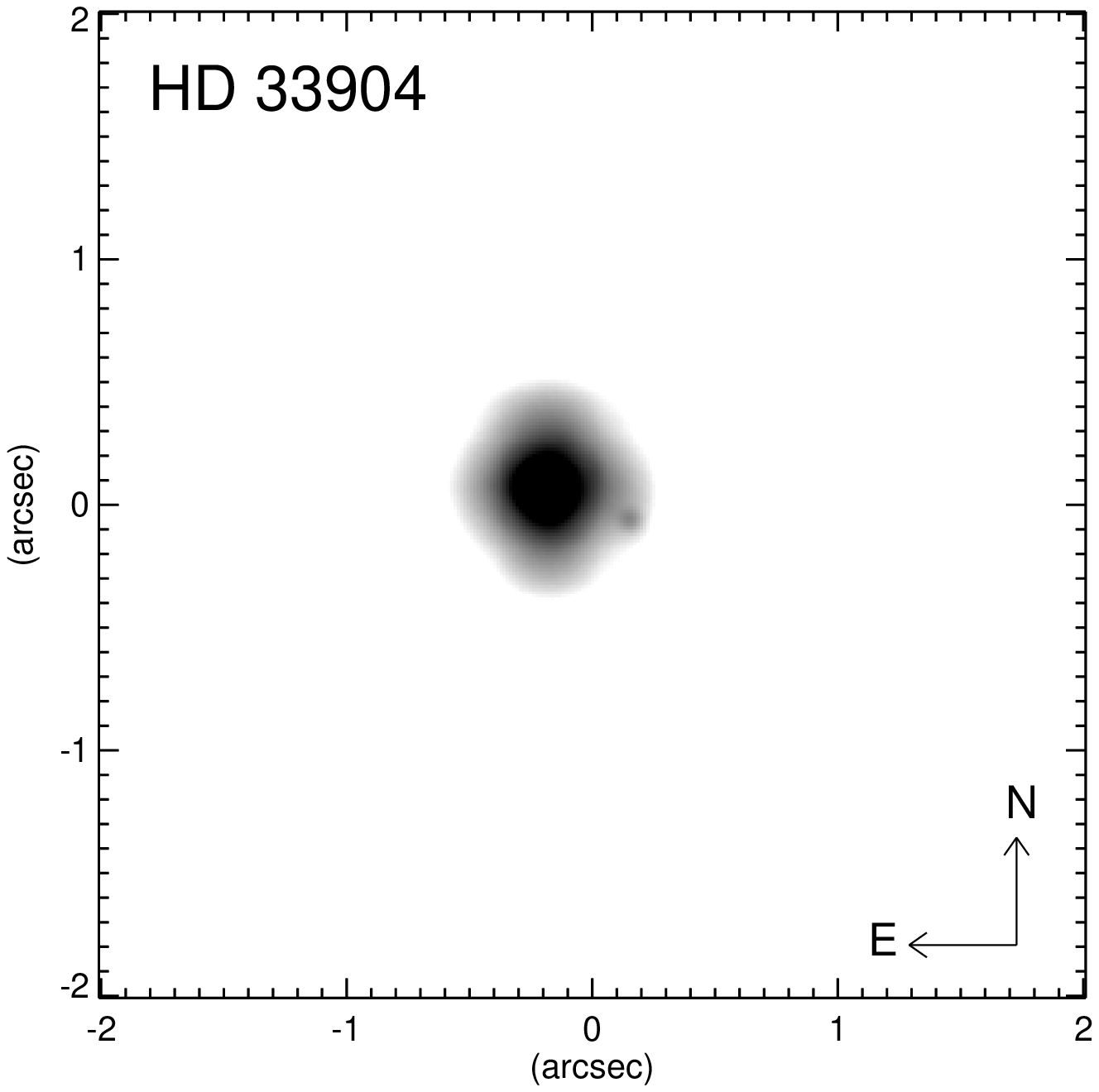}
\includegraphics[width=0.24\textwidth, angle=0]{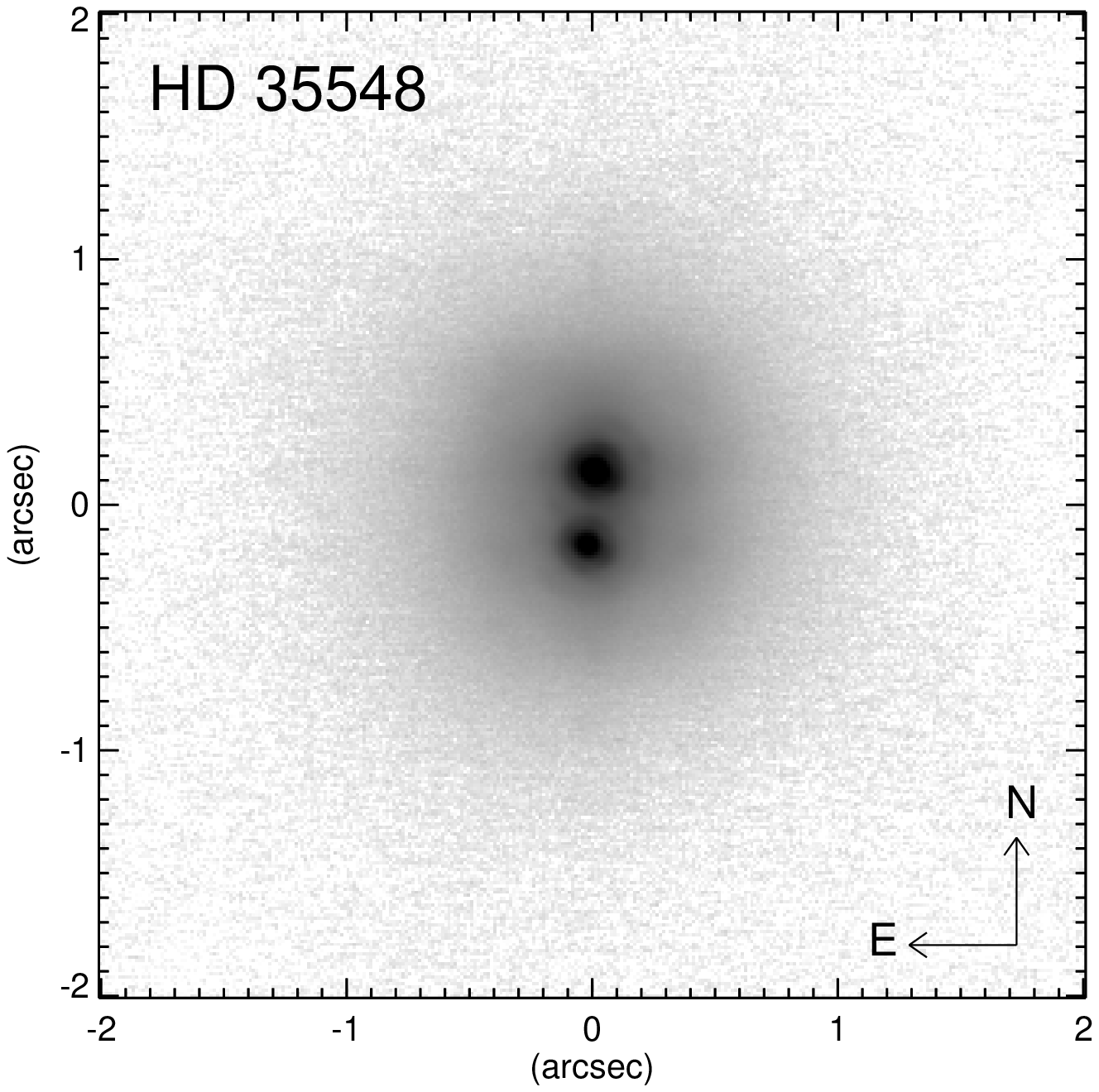}
\includegraphics[width=0.24\textwidth, angle=0]{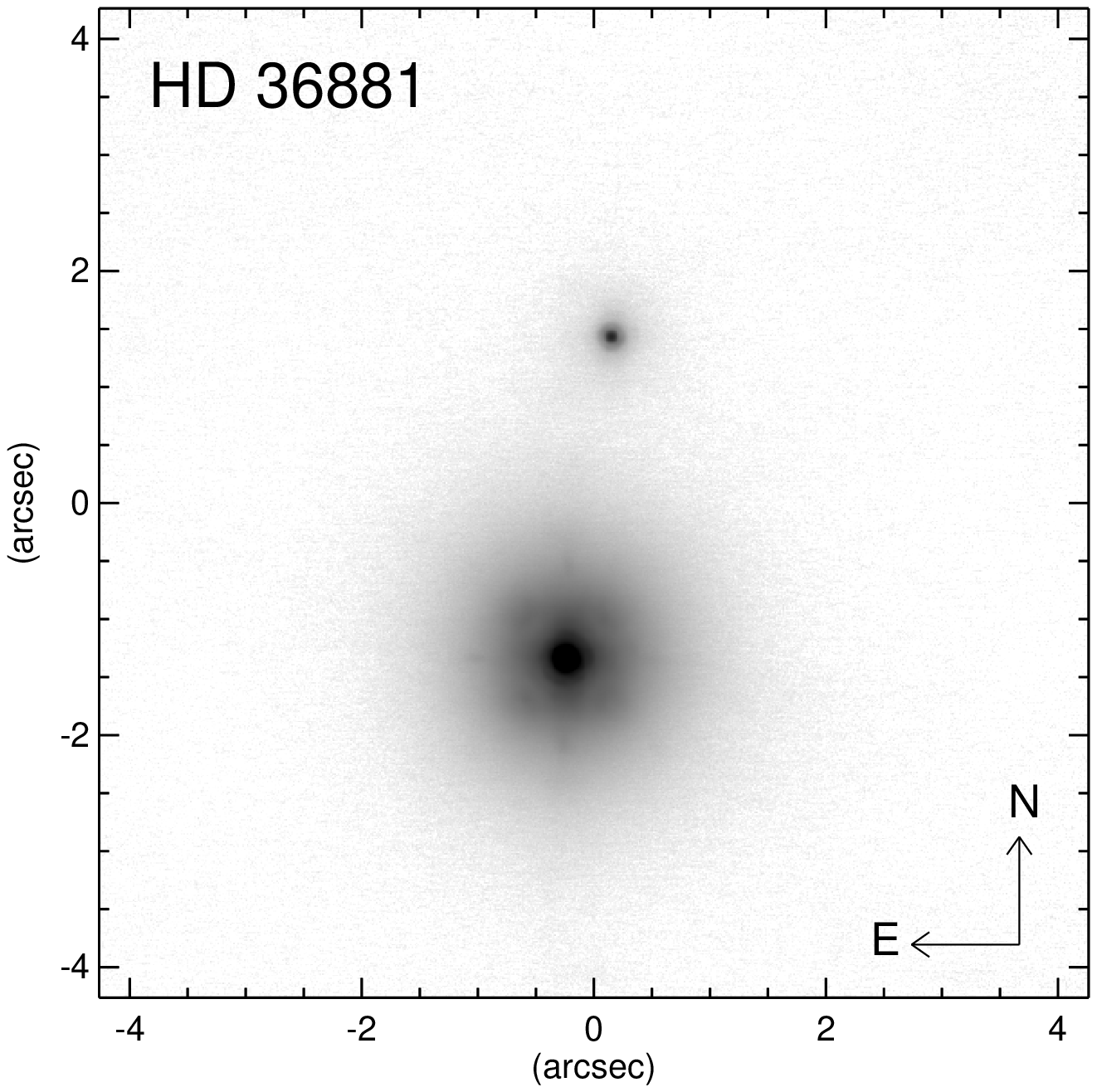}
\includegraphics[width=0.24\textwidth, angle=0]{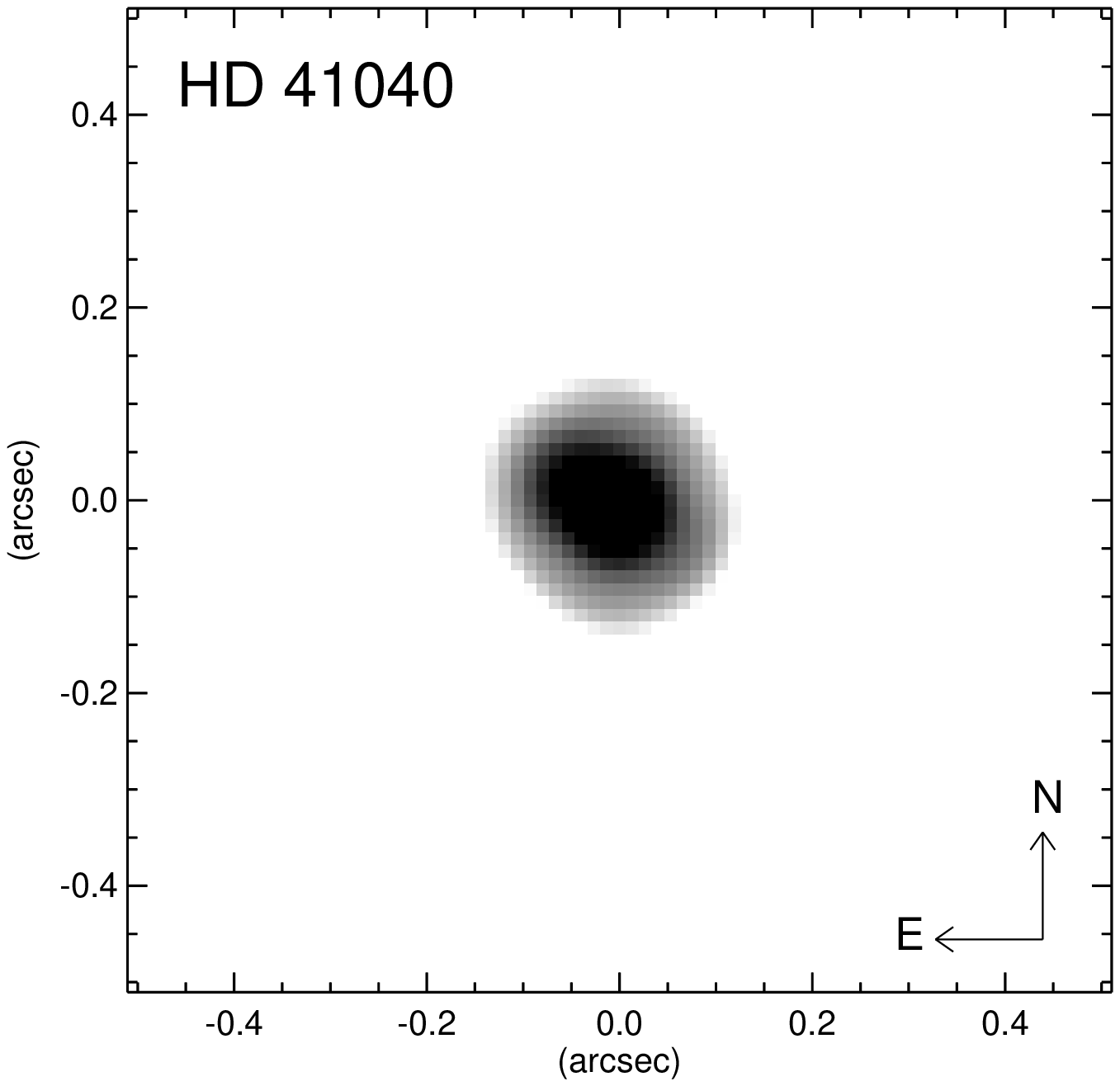}
\includegraphics[width=0.24\textwidth, angle=0]{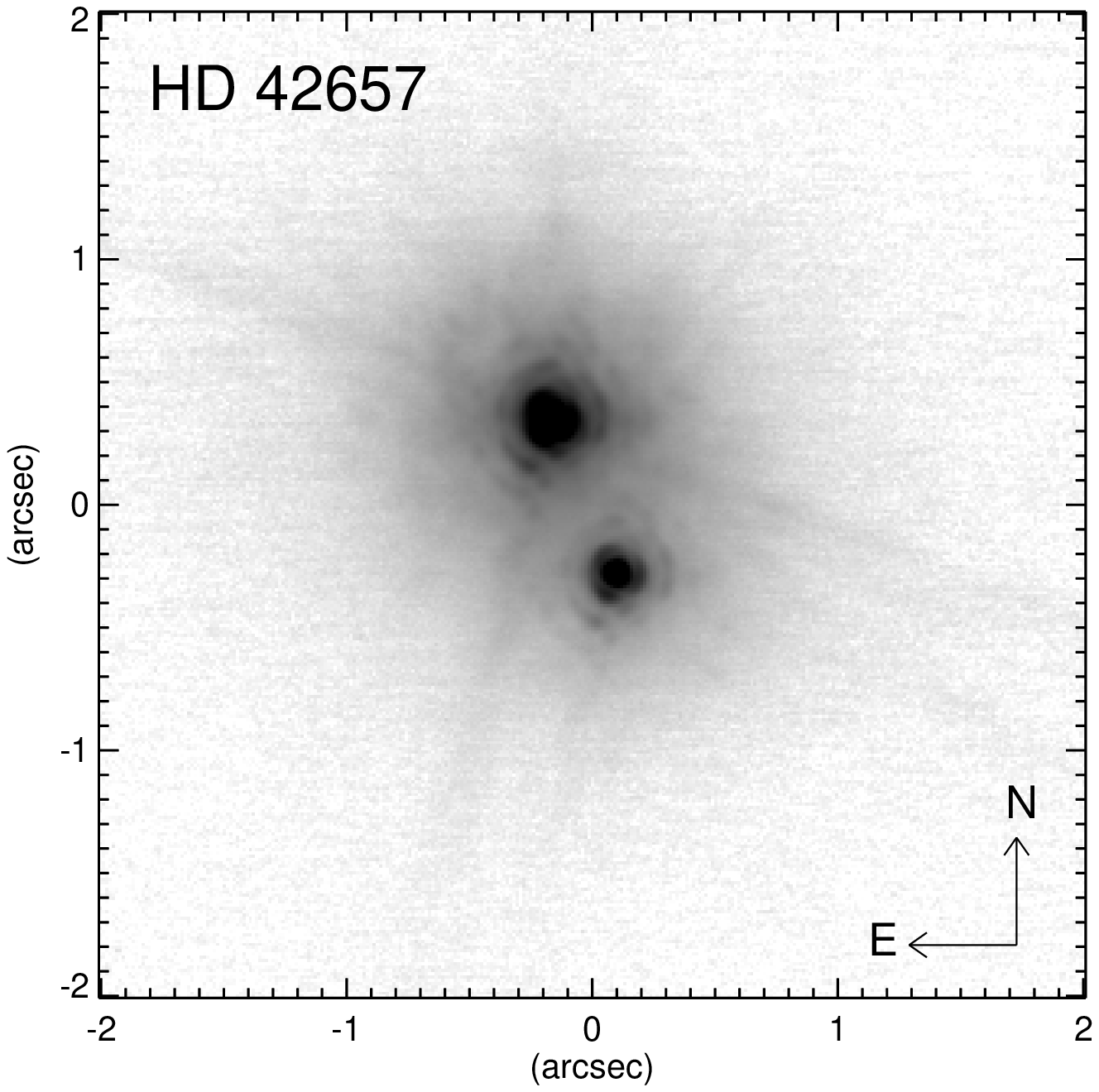}
\includegraphics[width=0.24\textwidth, angle=0]{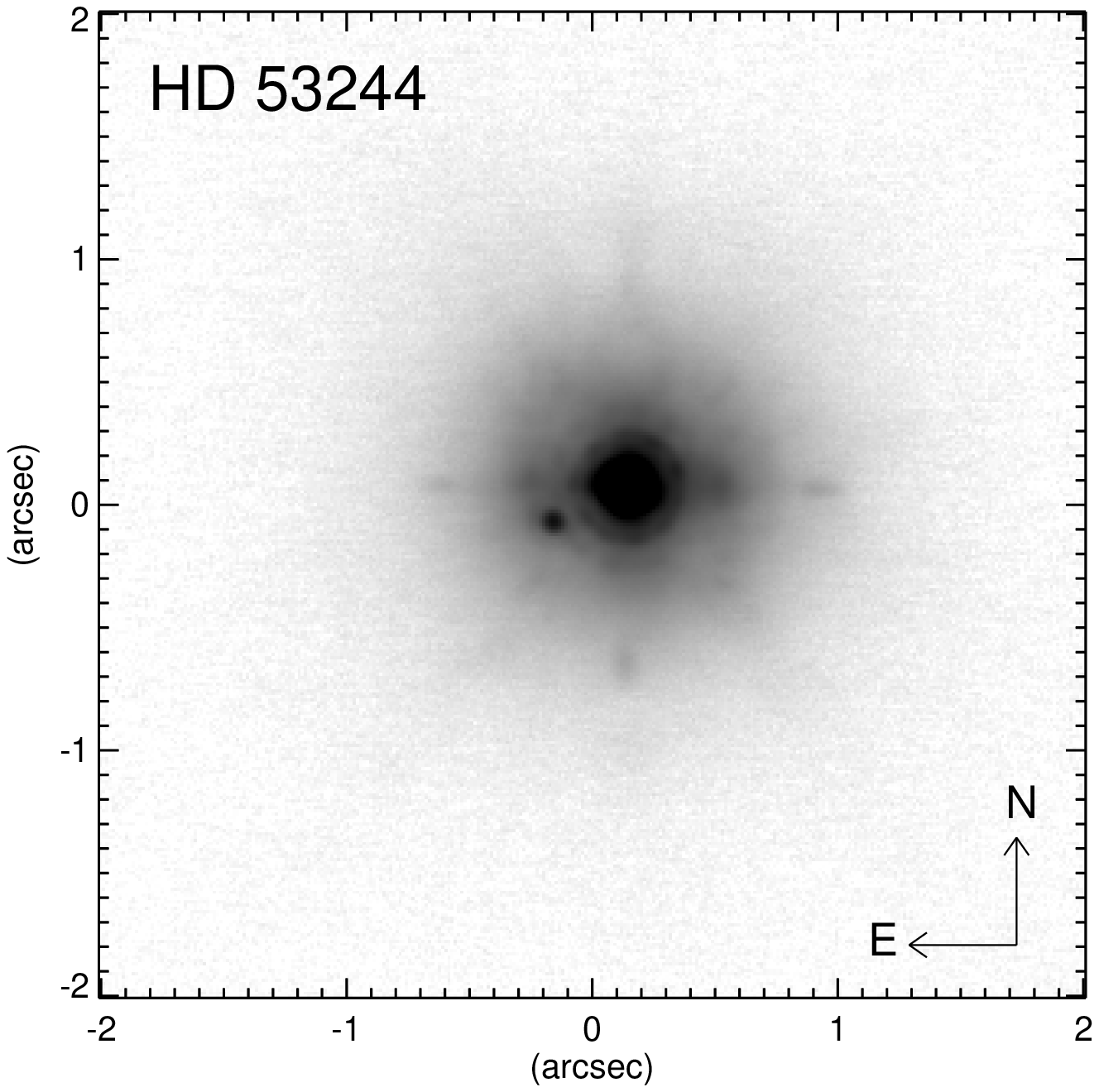}
\caption{
Images of the binaries detected in our VLT/NACO survey.
}
\label{fig:binaries}
\end{figure*}

\addtocounter{figure}{-1}

\begin{figure*}
\centering
\includegraphics[width=0.24\textwidth, angle=0]{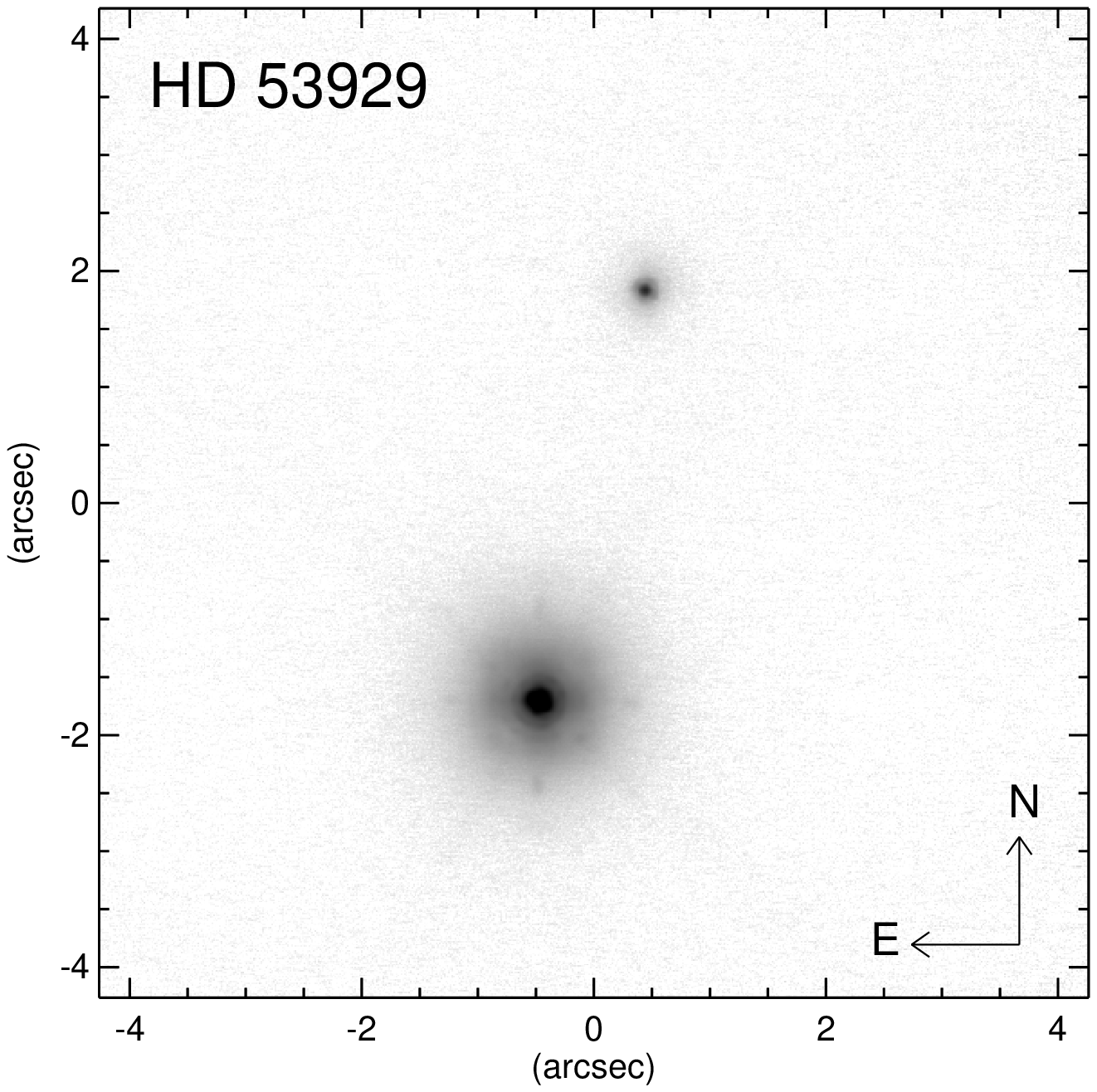}
\includegraphics[width=0.24\textwidth, angle=0]{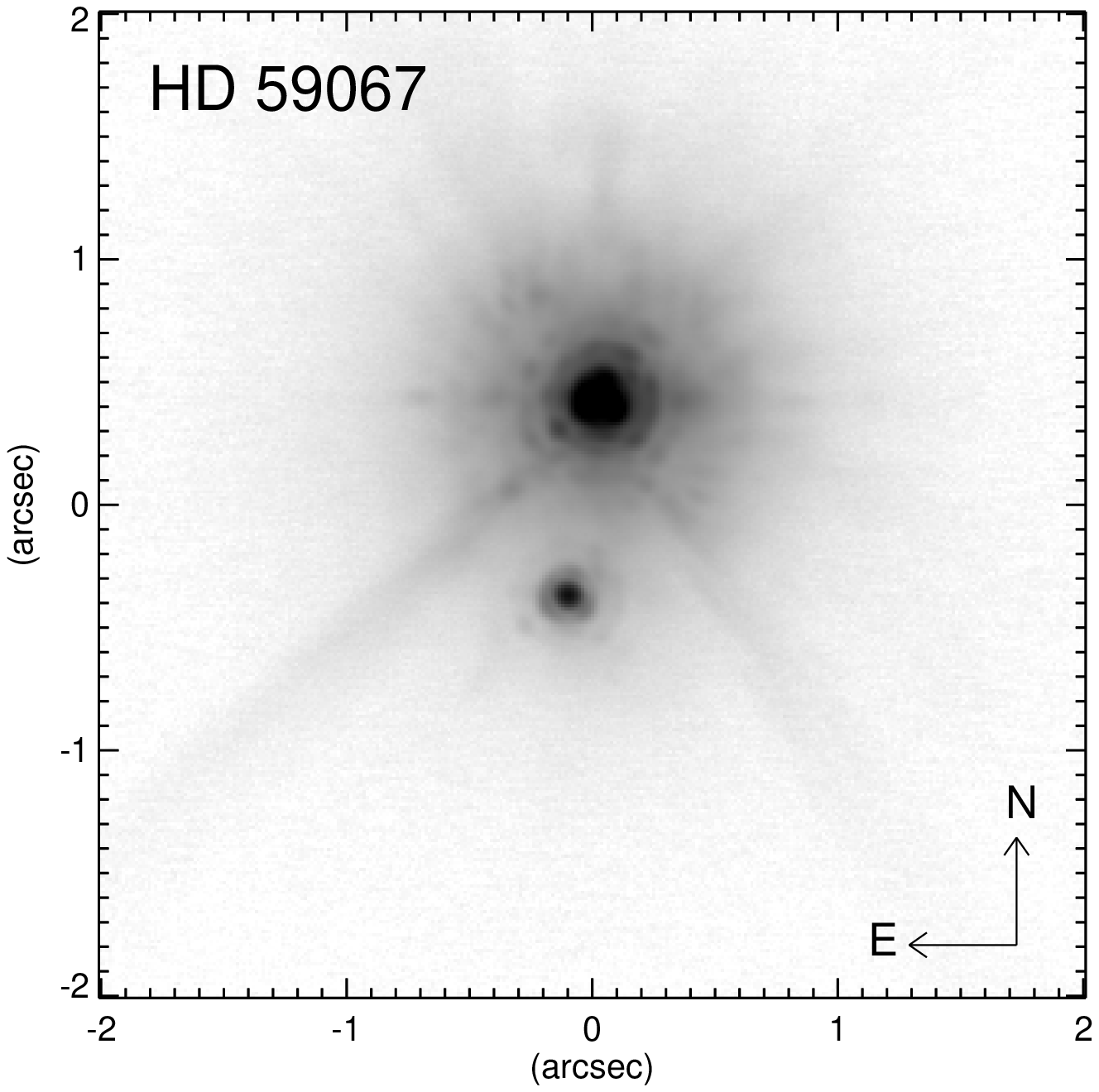}
\includegraphics[width=0.24\textwidth, angle=0]{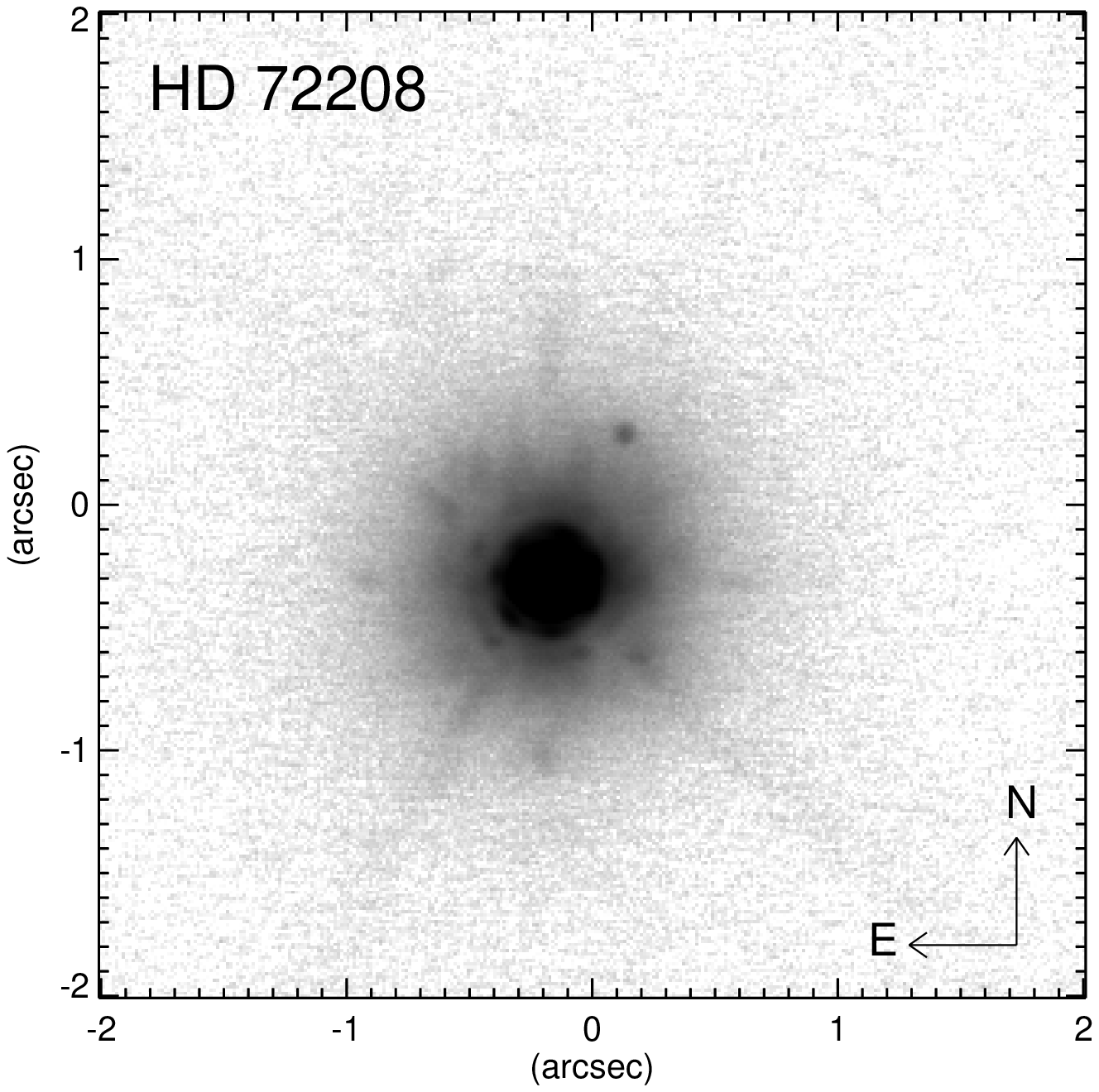}
\includegraphics[width=0.24\textwidth, angle=0]{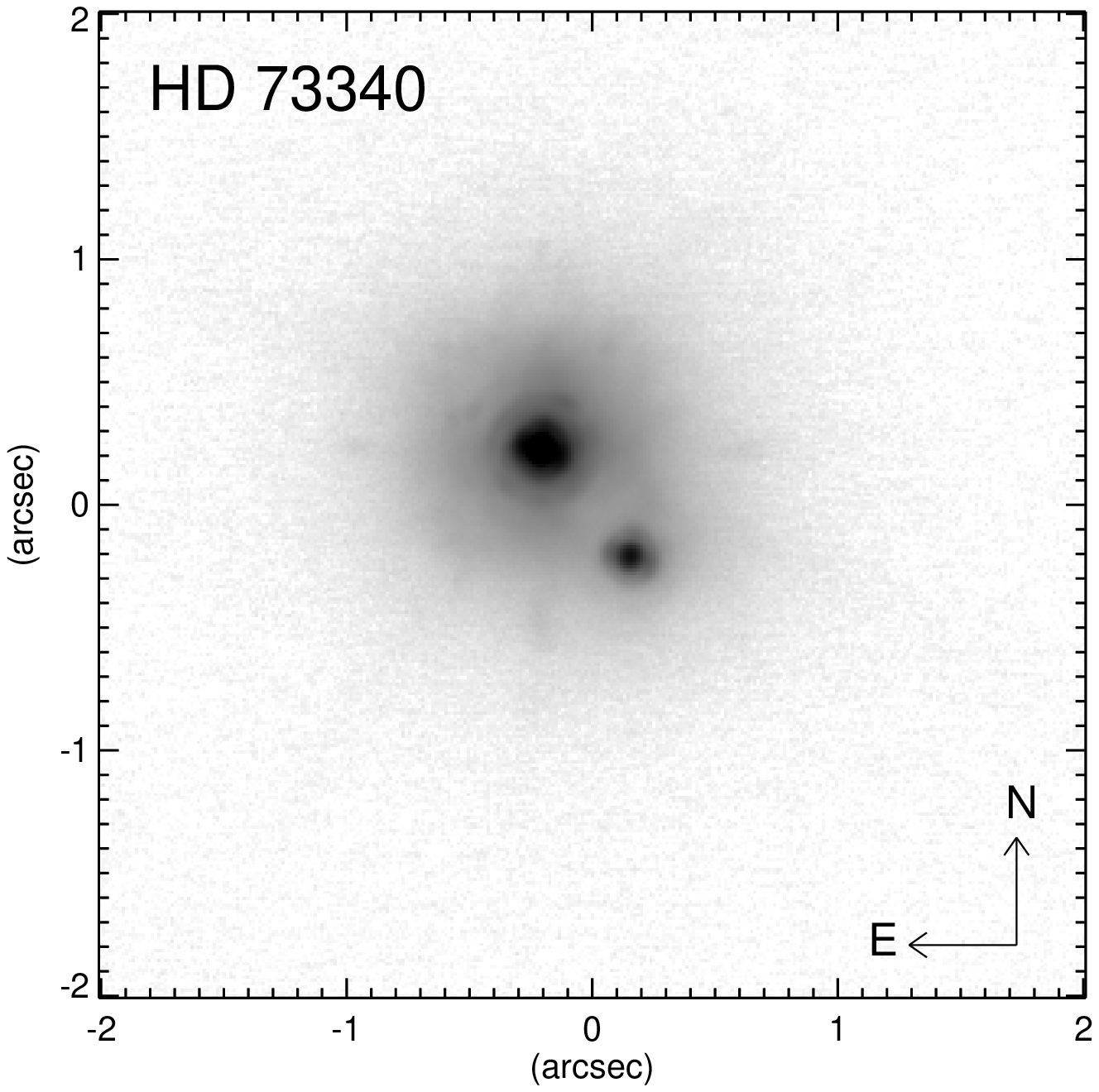}
\includegraphics[width=0.24\textwidth, angle=0]{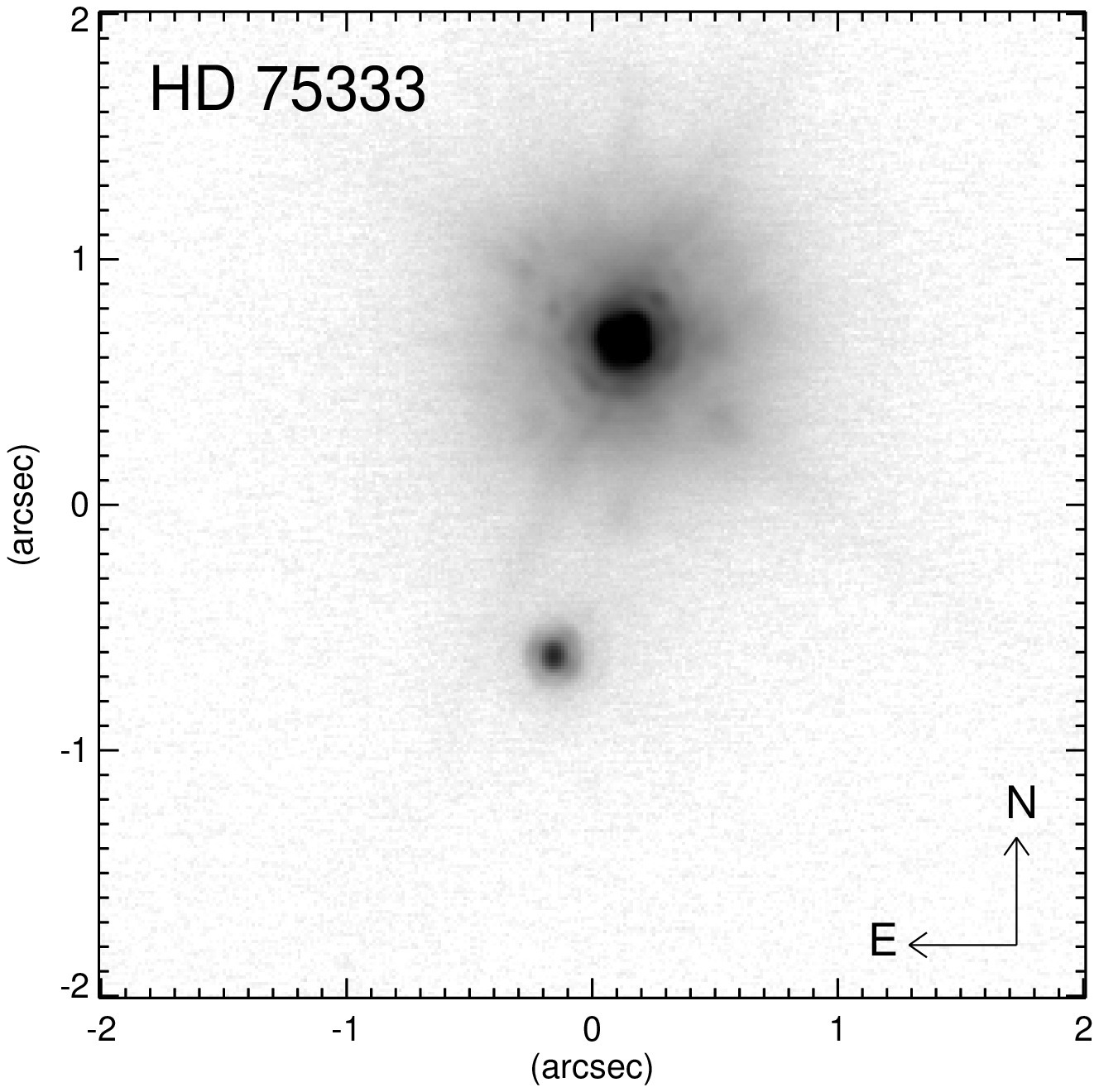}
\includegraphics[width=0.24\textwidth, angle=0]{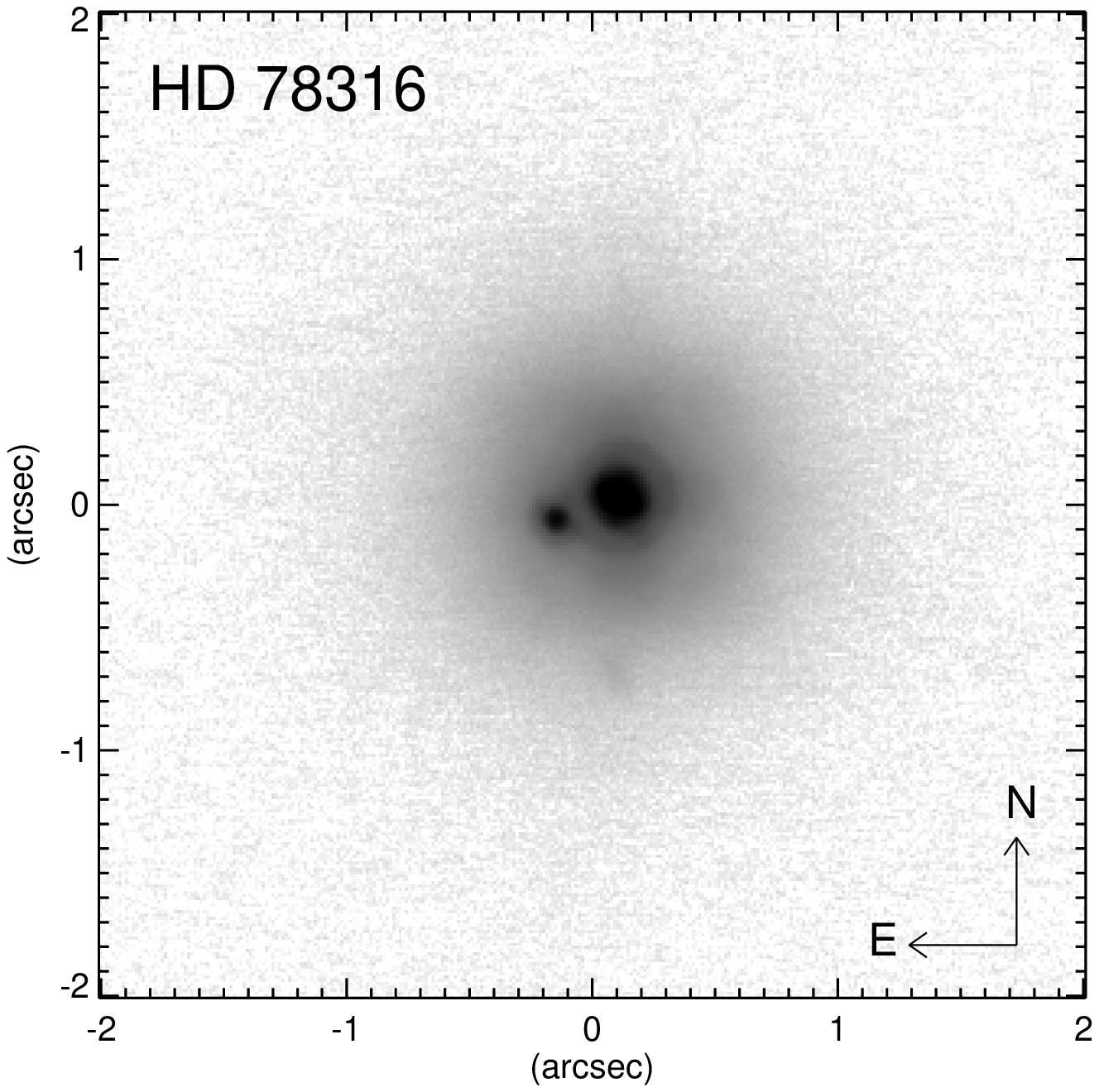}
\includegraphics[width=0.24\textwidth, angle=0]{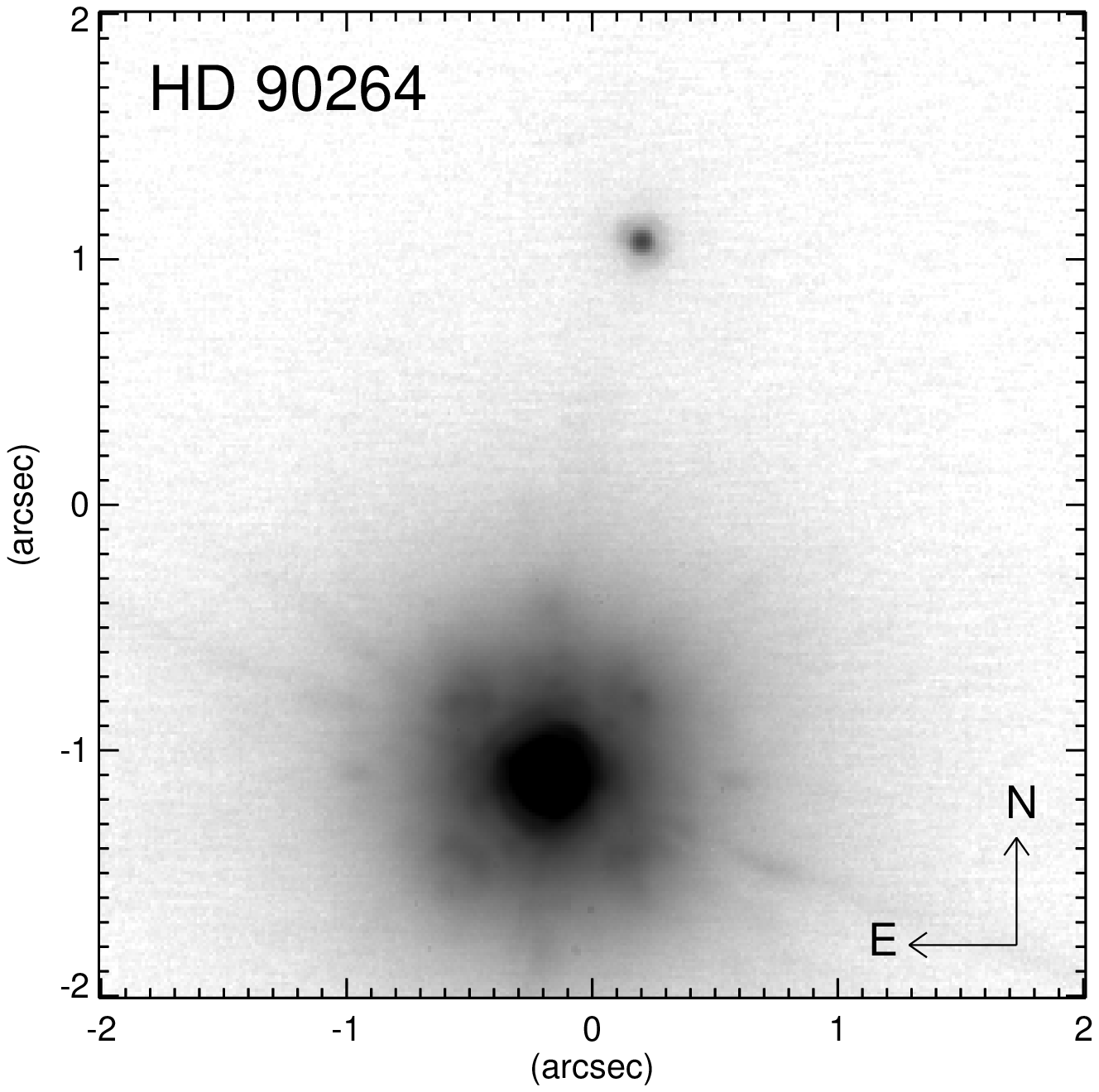}
\includegraphics[width=0.24\textwidth, angle=0]{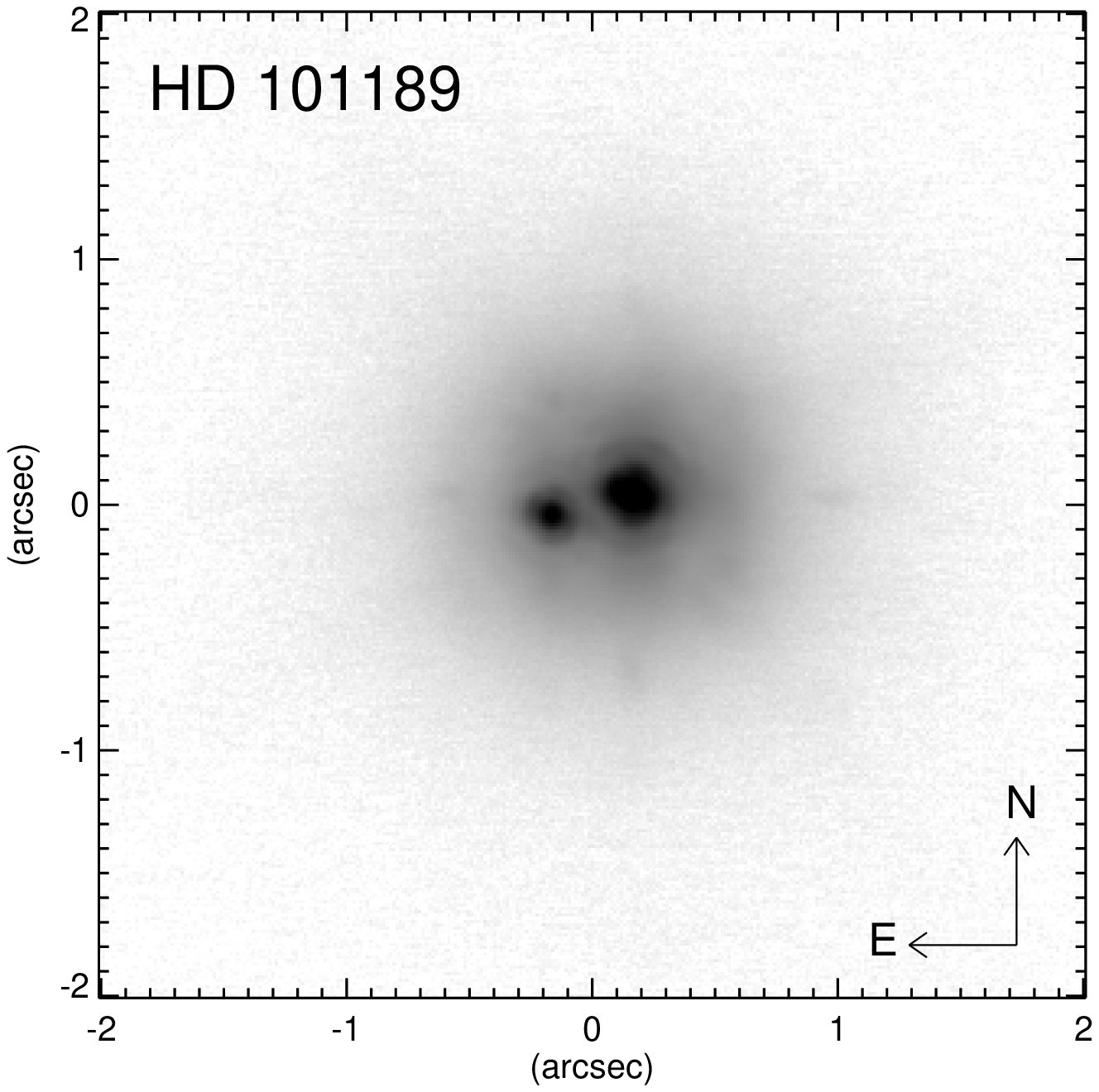}
\includegraphics[width=0.24\textwidth, angle=0]{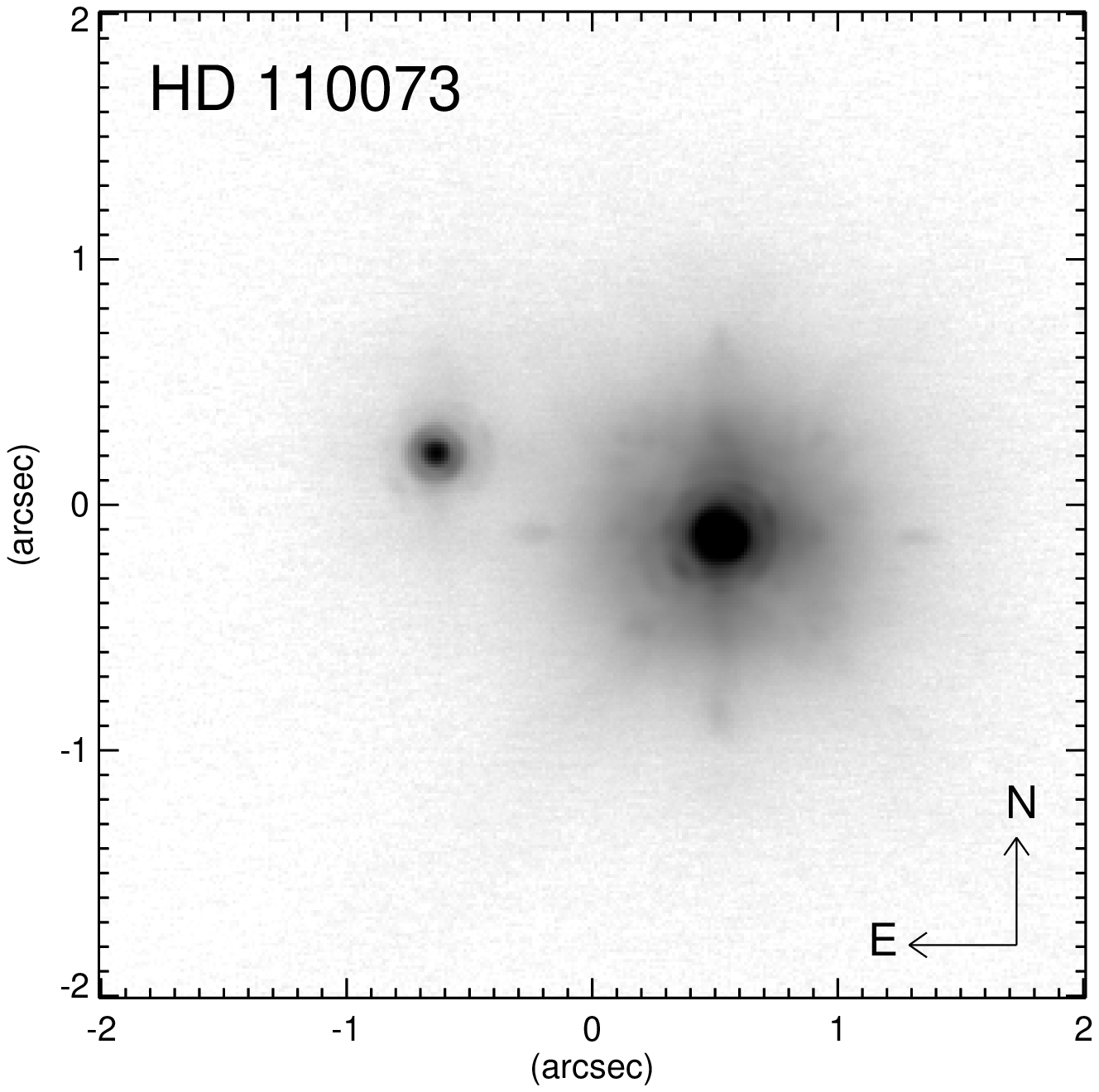}
\includegraphics[width=0.24\textwidth, angle=0]{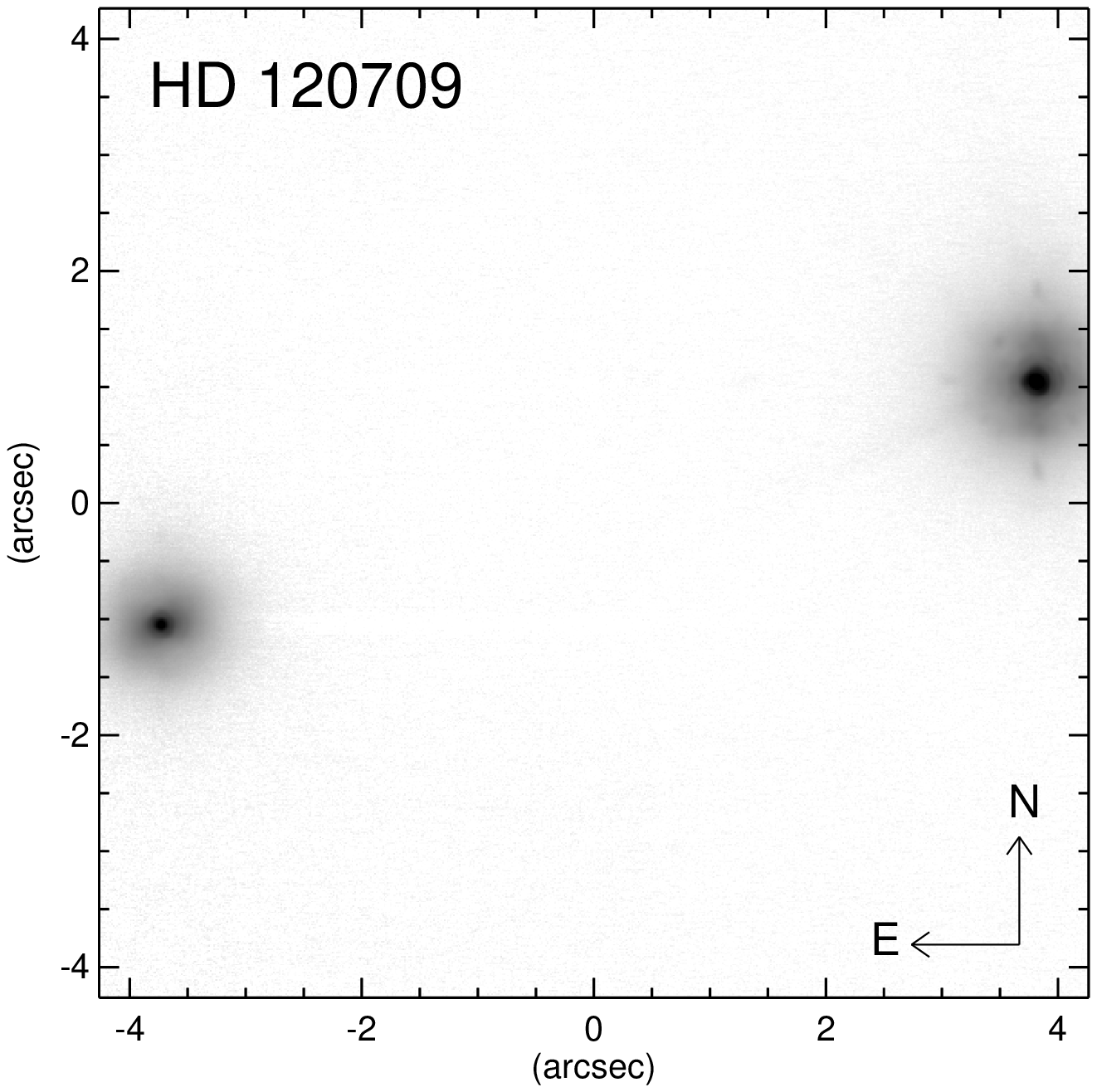}
\includegraphics[width=0.24\textwidth, angle=0]{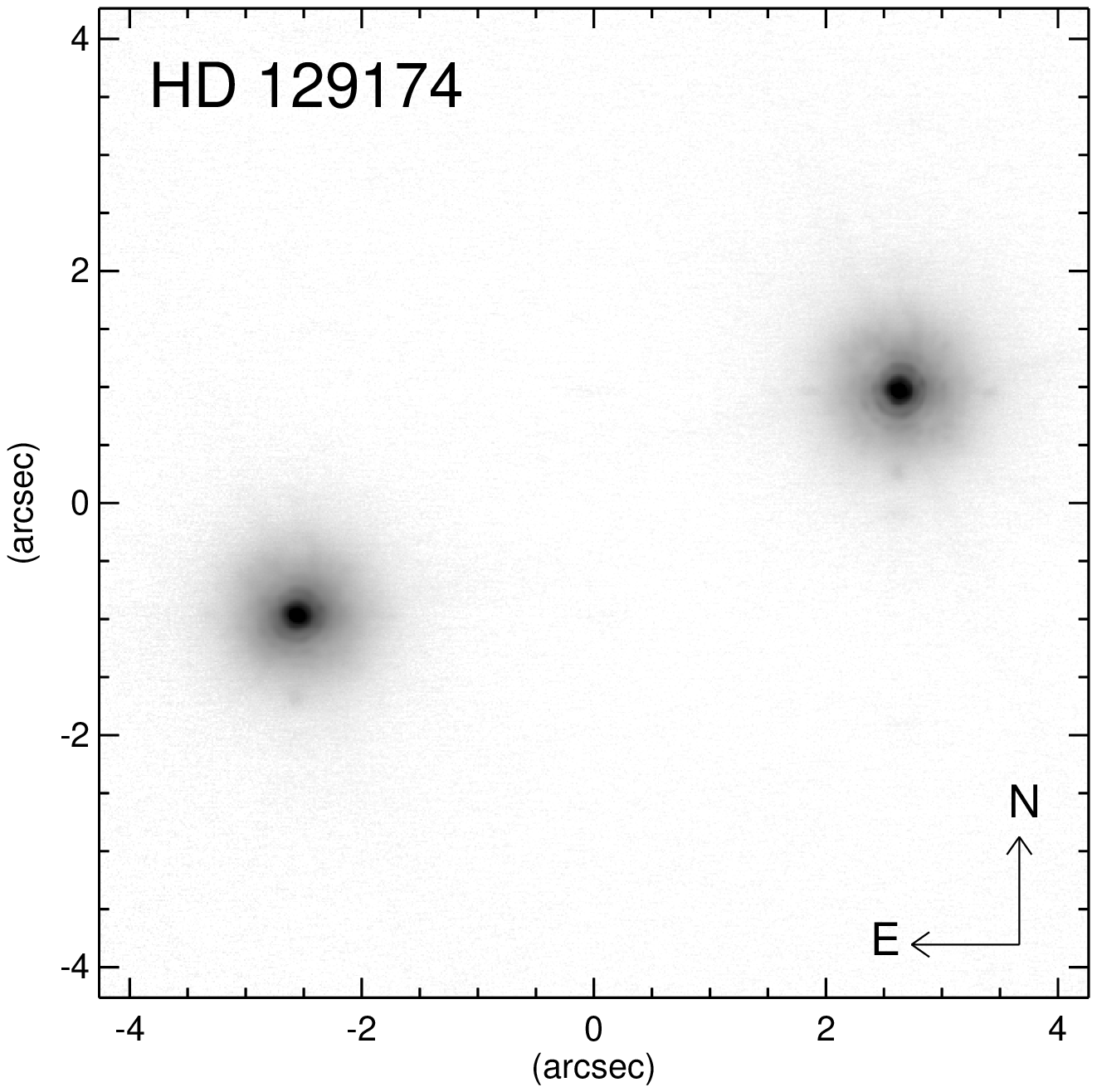}
\includegraphics[width=0.24\textwidth, angle=0]{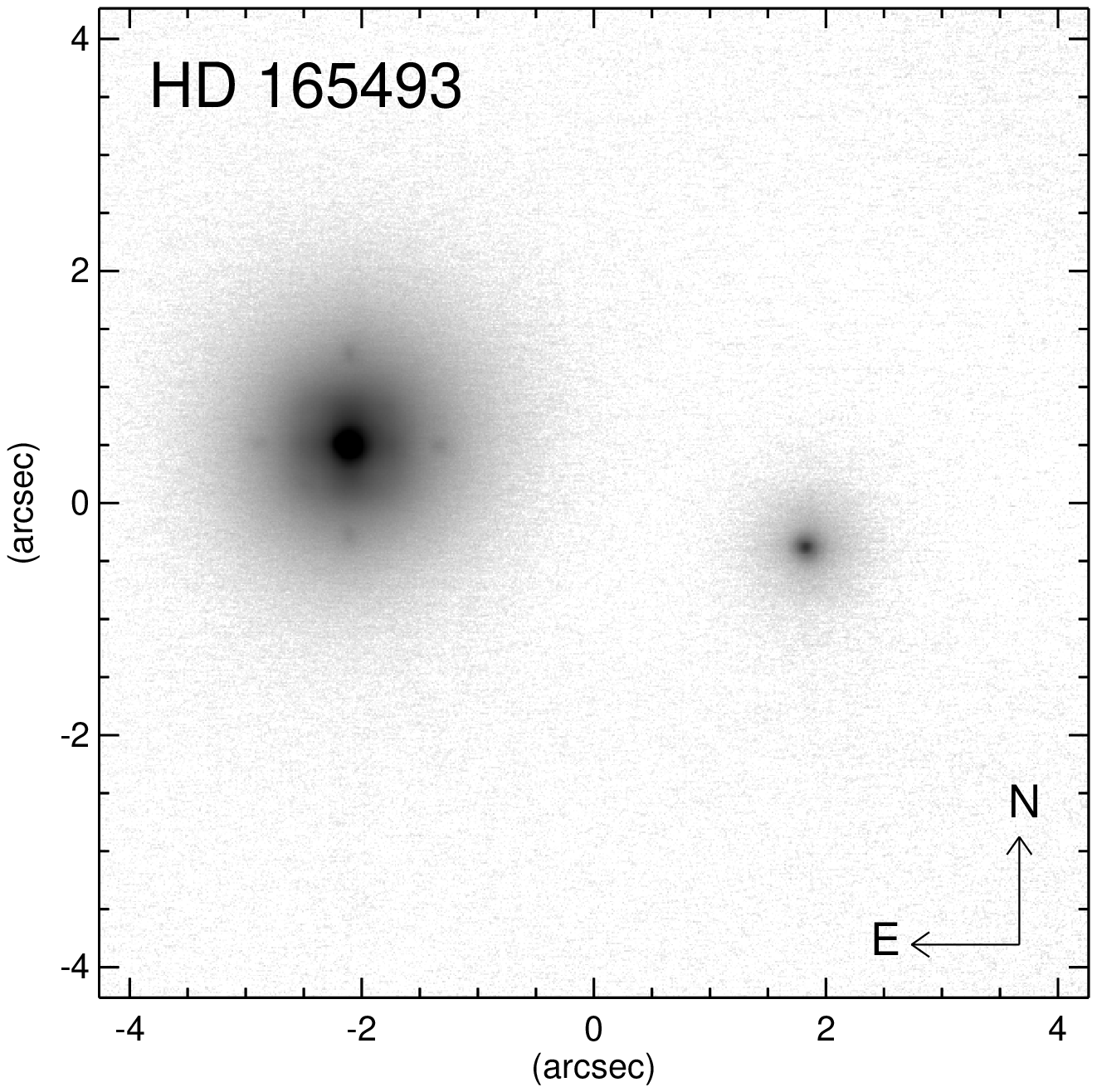}
\includegraphics[width=0.24\textwidth, angle=0]{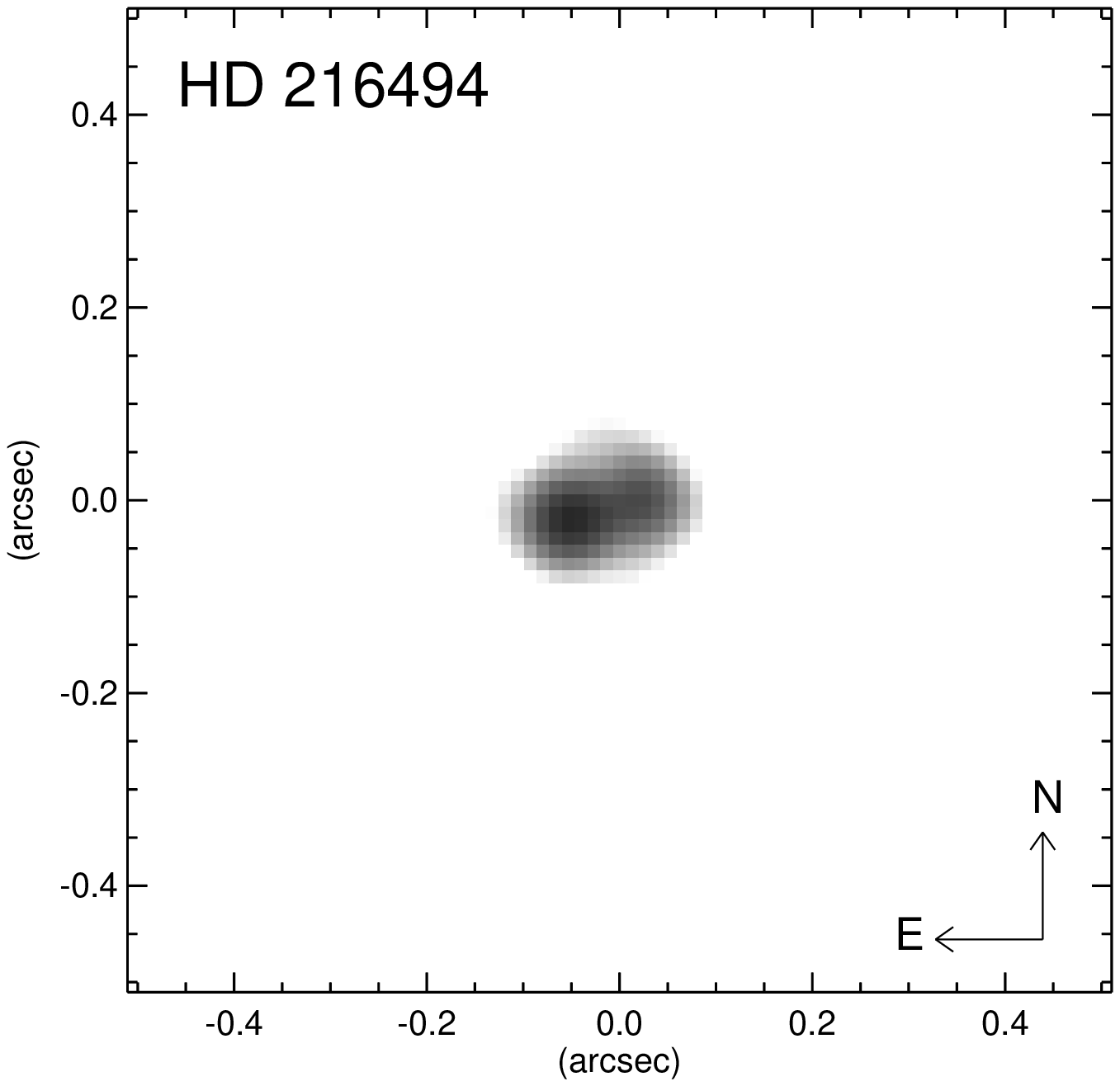}
\includegraphics[width=0.24\textwidth, angle=0]{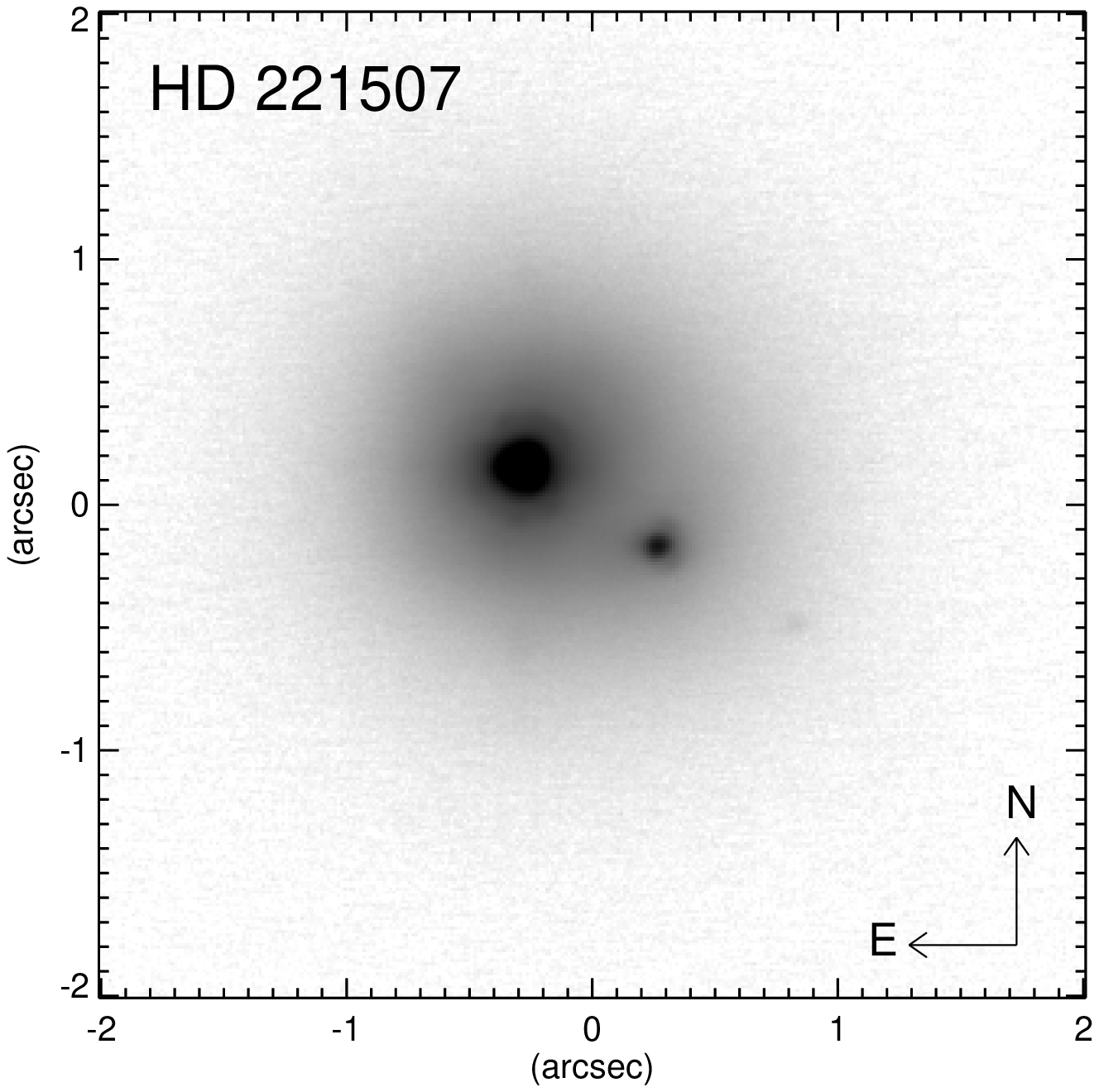}
\caption{
(continued)
Images of the binaries detected in our VLT/NACO survey.
}
\end{figure*}

\begin{figure*}
\centering
\includegraphics[width=0.32\textwidth, angle=0]{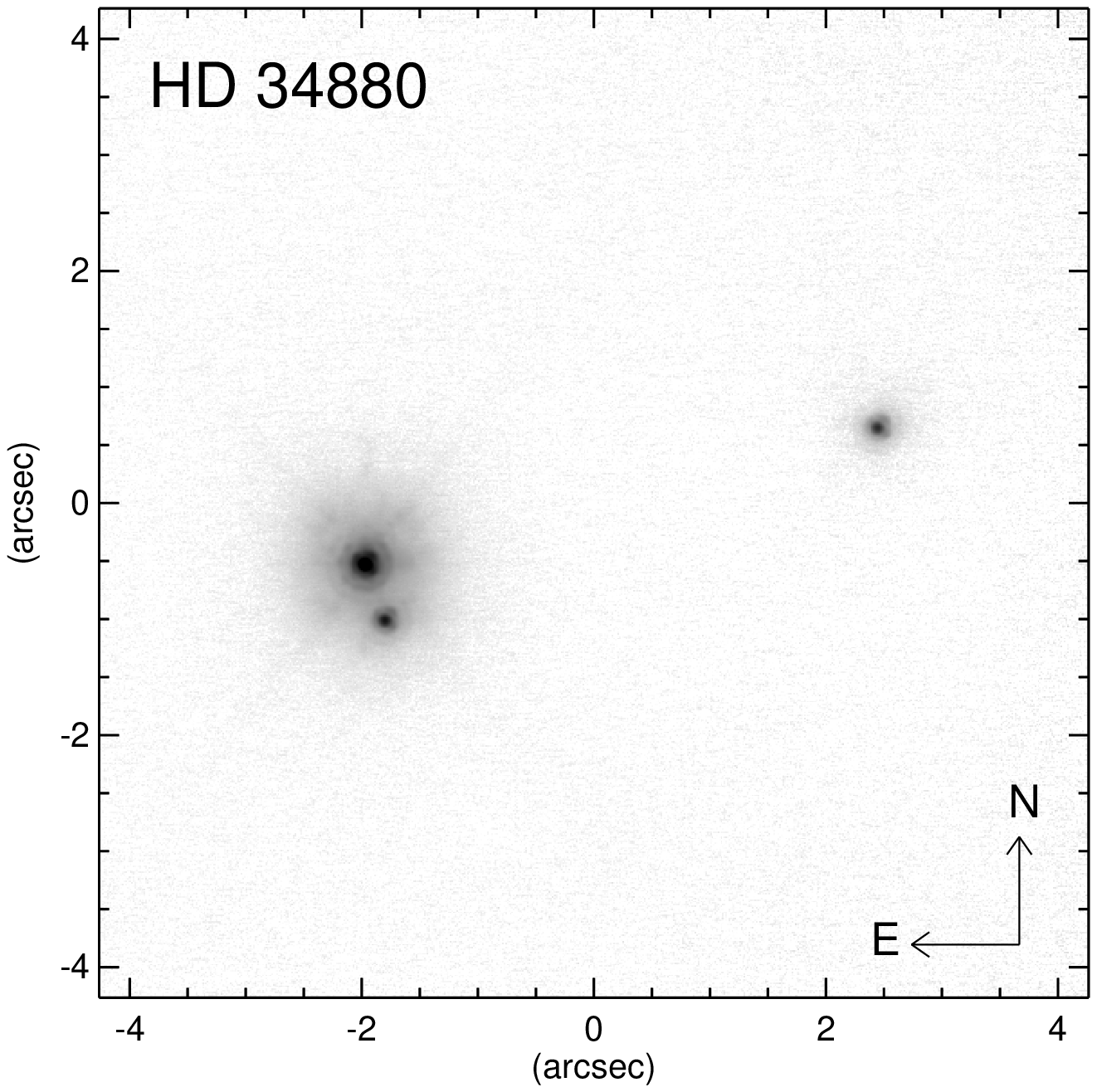}
\includegraphics[width=0.32\textwidth, angle=0]{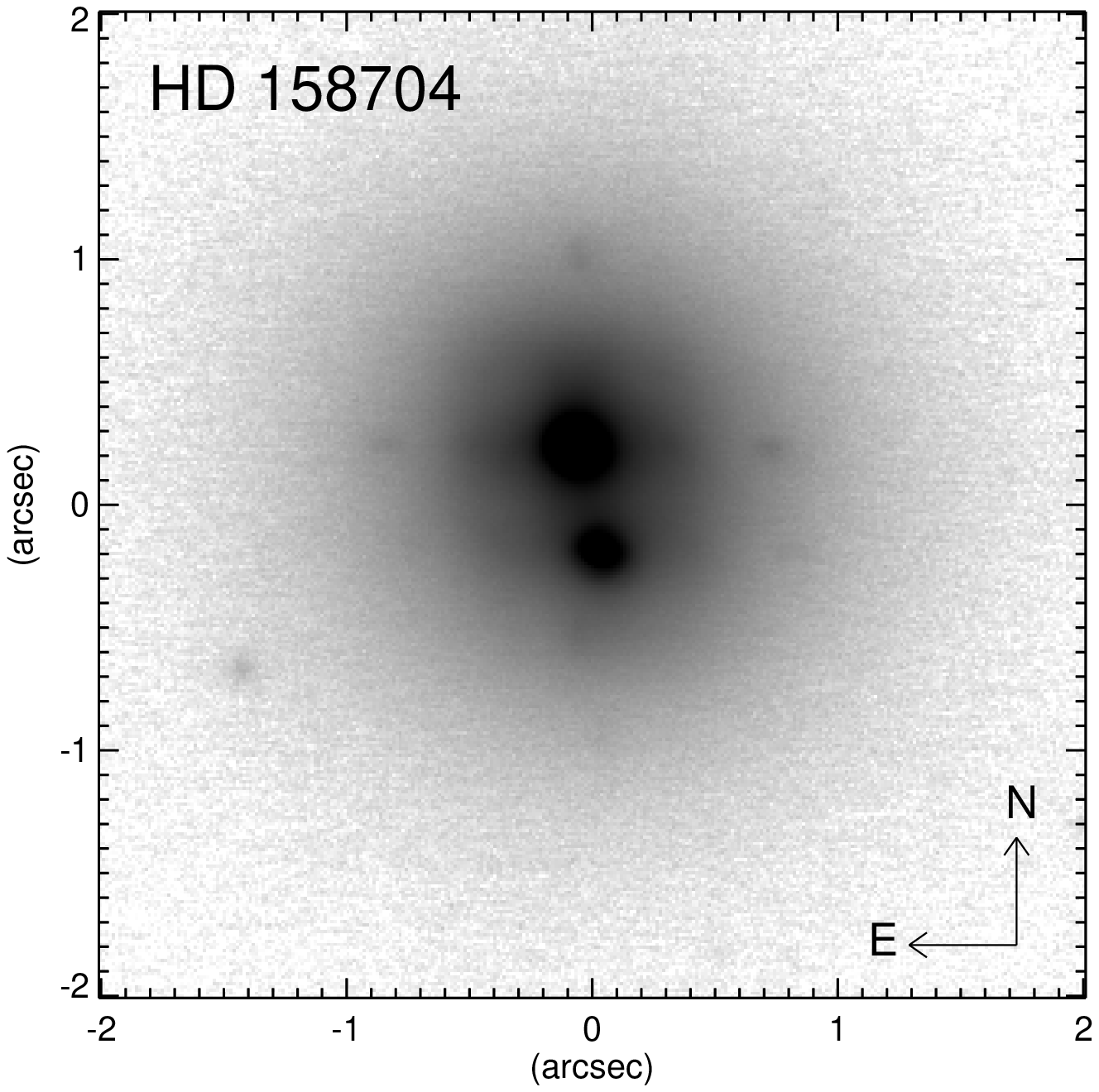}
\includegraphics[width=0.32\textwidth, angle=0]{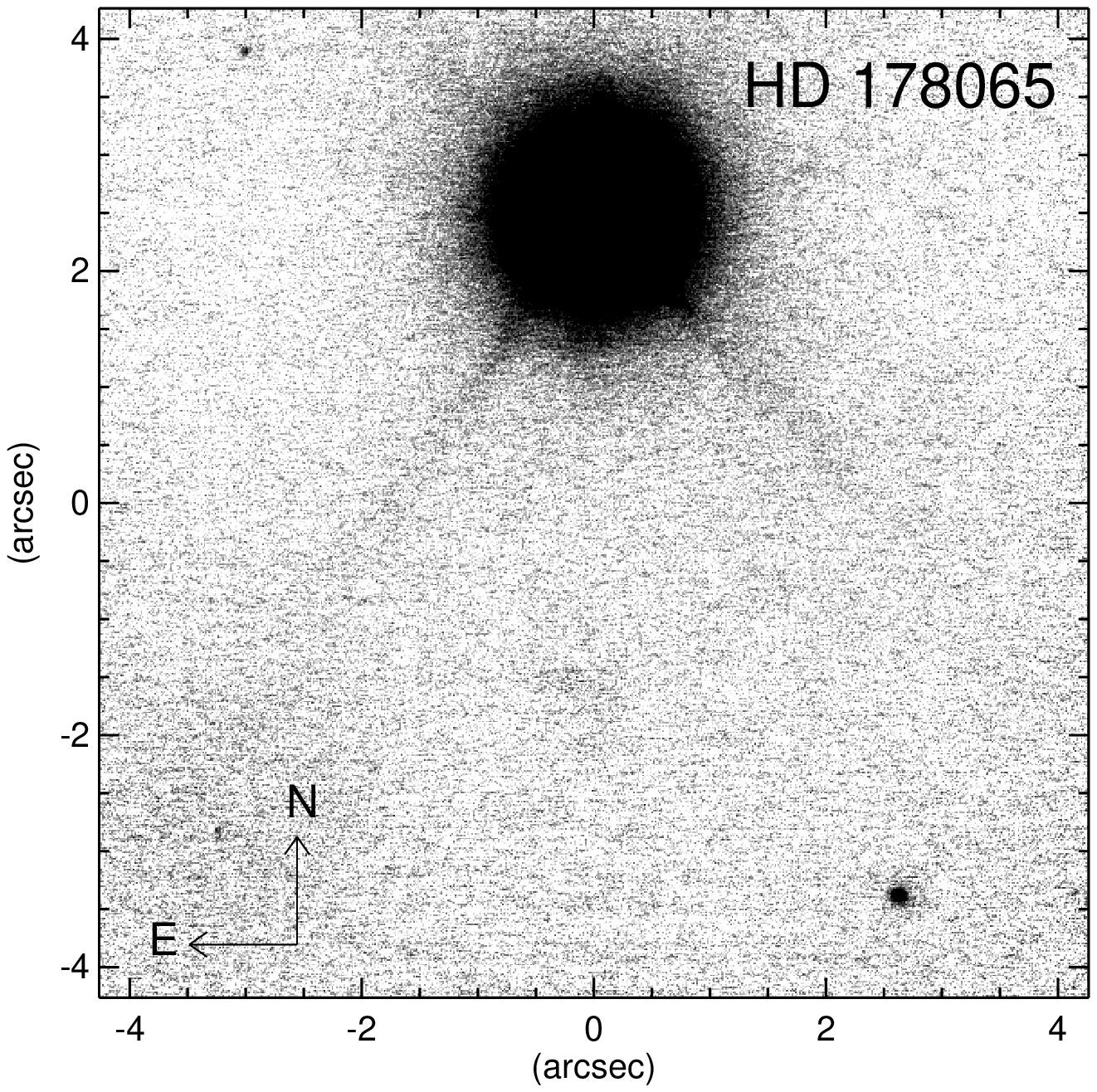}
\caption{
Images of the triple systems detected in our VLT/NACO survey.
}
\label{fig:triples}
\end{figure*}

\begin{figure}
\centering
\includegraphics[width=0.40\textwidth, angle=0]{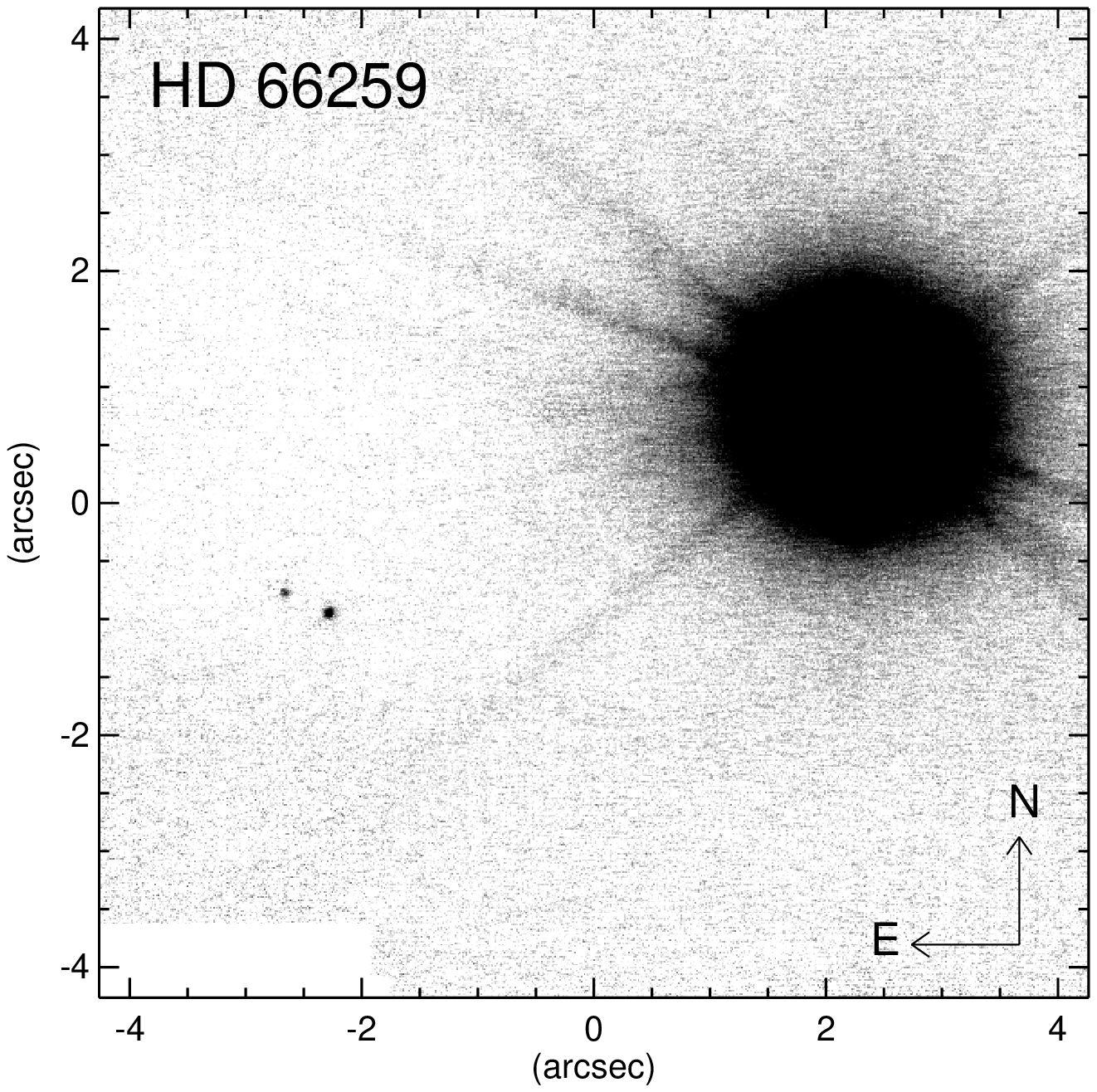}
\includegraphics[width=0.40\textwidth, angle=0]{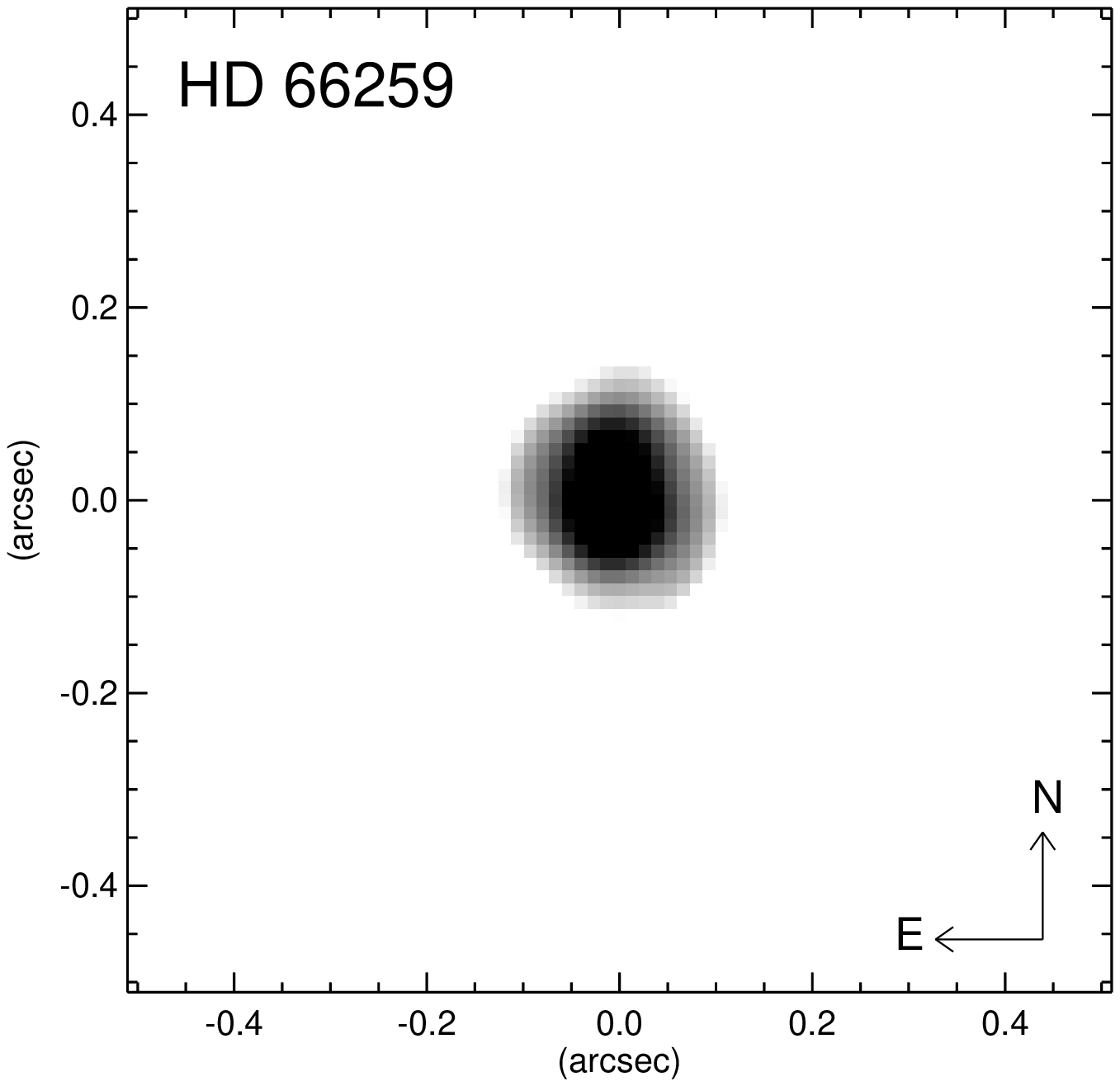}
\caption{
Images of the wide pair (top) and the close pair (bottom) of the quadruple system detected in our VLT/NACO survey.
}
\label{fig:quadruples}
\end{figure}

The images of the resolved binaries are shown in the two parts of Fig.~\ref{fig:binaries}. 
The triple systems can be found in Fig.~\ref{fig:triples} and the quadruple system in Fig.~\ref{fig:quadruples}.
All images are displayed using a logarithmic scale.
In the images with the closest companion candidates, showing just the inner 1\arcsec{}, this logarithmic scale
had to be adapted to enhance the image details.
The same modification was applied to HD\,33904.

\subsection{Limits for undetected companions and completeness}
\label{sect:completeness}

\begin{figure}
\centering
\includegraphics[width=0.45\textwidth, angle=0]{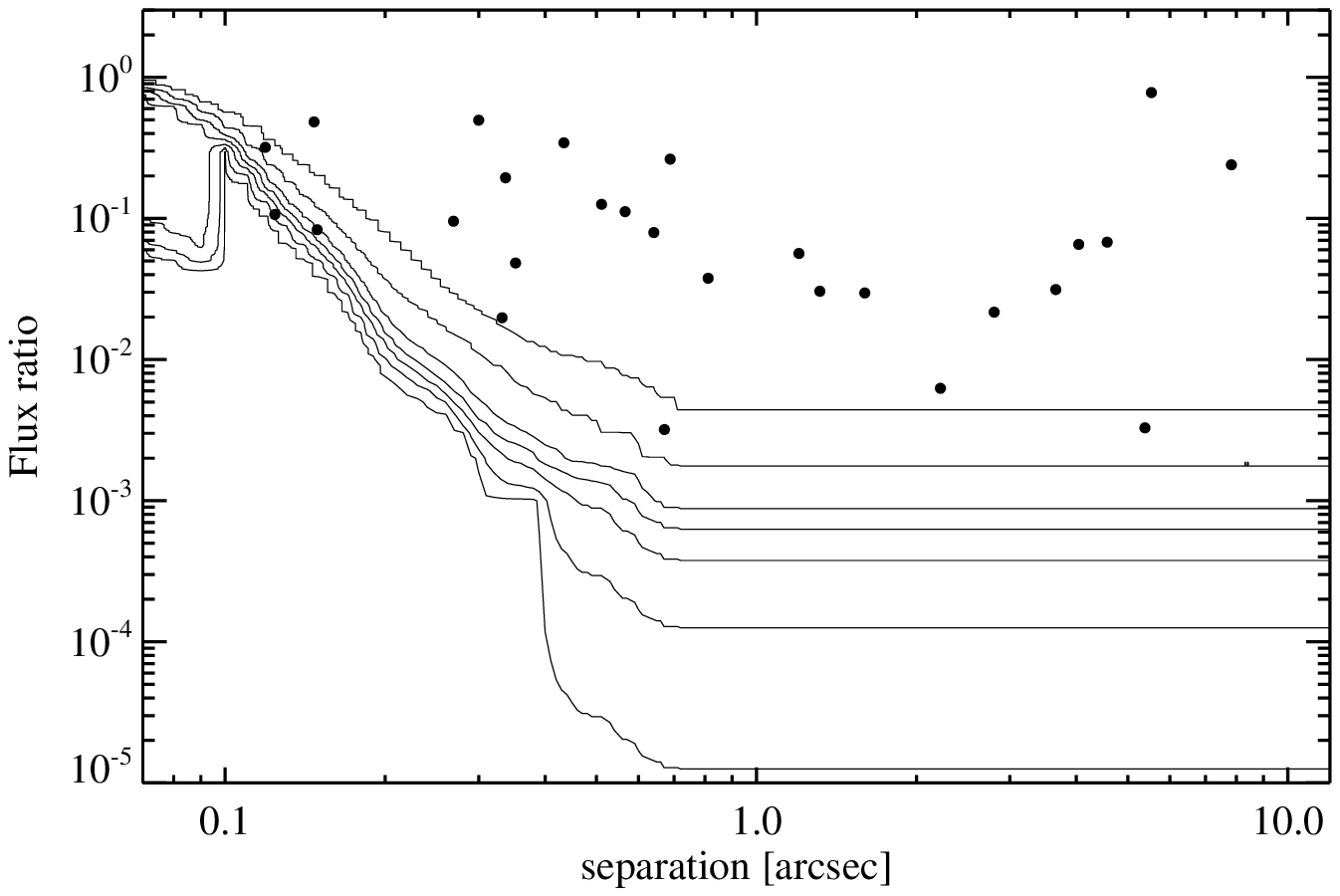}
\caption{
Completeness map for our survey.
The lines represent the completeness of our observations as derived
from the sensitivity limit for undetected companions (see Sect.~\ref{sect:completeness}).
The lines are, from top to bottom, 99\%, 90\%, 70\%, 50\%, 30\%, 10\%, and 1\% completeness.
The circles represent the binaries found in our study.
}
\label{fig:compl_maps}
\end{figure}

The detection limits were computed using the method described in Correia et al.\ (\cite{Correia2006}).
At each radial distance and position angle from the star, the standard deviation in the flux was 
calculated over a circular region of radius 70\,mas, i.e., equivalent 
to the mean size of the point spread function (PSF) core.
The detection limit as a function of separation from the star
is the average of the 5$\sigma$ flux over 
all position angles except those lying in the direction of the companion candidate.

For any given component of the binaries as well as for the unresolved sources,
we thus have the limiting flux ratio for undetected companions as a function of separation.
We can therefore produce a completeness map for 
each of these sources, i.e.\ a map giving the probability of detecting
a companion as a function of separation and magnitude difference.
The total completeness map of all 
sources is simply the average of all individual completeness maps.

Figure~\ref{fig:compl_maps} shows the total completeness map.
The lines represent the completeness of our observations at levels
99\%, 90\%, 70\%, 50\%, 30\%, 10\%, and 1\% completeness,
from top to bottom.
The circles represent the binaries found in our study.
The majority of the companion candidates falls above the 
99\% completeness level
and three more fall above the 90\% completeness level.
The two sources below 70\% are very close to the first Airy ring,
where our method is probably very inaccurate.
Overall, we are confident in the completeness levels. 
We note that we plotted only objects with chance projection probabilities
below 1\% and do not show the objects with separations below 0\farcs1.
Assuming that both the distribution of separation and the 
distribution of flux ratio of companions are flat (which is obviously
a very rough assumption),
we estimate that the completeness is above 90\% for separations of
a) between 0\farcs2 and 0\farcs4 and above a flux ratio of 10$^{-1}$,
b) between 0\farcs4 and 0\farcs7 and above a flux ratio of 10$^{-2}$,
and c) between 0\farcs7 and 8\arcsec{} and above a flux ratio of 2$\times$10$^{-3}$.
It should be noted that the S13 camera of NACO does not allow to detect companions 
at large separations ($>\sim$7--8\arcsec{}).

For HD\,41040 and HD\,66259, we claim companion candidates within the diffraction limit, which
are essentially elongations of the PSF.
In both cases, we carefully analyzed not only the combined images, but also the
individual frames, and were convinced that the results are real.
For HD\,66259, this is supported by the two PSFs of the other companion candidates
in the image not displaying this elongation.

For HD\,29589 and HD\,31373, we find hints for companions directly on the
Airy ring, which we consider as artifacts.

\subsection{Chance projections}
\label{sect:projections}

To identify the systems whose components are 
gravitationally bound and those that are only the result of a
chance projection, we used a statistical approach
(see e.g.\ Correia et al.\ \cite{Correia2006}).
In a first step, we determined the local surface density of background/foreground sources in each field.
For this purpose, we compiled the number of 2MASS objects brighter
than the companion candidates in the K-band in a 30\arcmin{}$\times$30\arcmin{}
field surrounding each primary.
This leads to the average surface density of objects brighter than the limiting 
magnitude $\Sigma(K < K_{\rm comp})$.
Assuming a random uniform 
distribution of unrelated objects across the field, the resulting 
probability $P(\Sigma, \Theta)$ of at least one unrelated source being located 
within a certain angular distance $\Theta$ from a particular target is 
given by 

\begin{displaymath}
P(\Sigma, \Theta) = 1 - e^{-\pi\Sigma\Theta^2}.
\end{displaymath}

The last column of Table~\ref{tab:astrometry} gives the resulting probability for a companion candidate to be unrelated
to the primary of a system.
Since the 2MASS Point Source Catalog is incomplete for stars fainter than
K=14.3, the calculated chance projection probabilities are only lower limits
for sources fainter than K=14.3.
All companion candidates detected in our survey in binaries have probabilities to be projected unrelated stars well below the percent level.
This means that they are very likely bound to their systems, although considering probabilities of individual sources is known to be prone 
to error (see e.g.\ Brandner et al.\ \cite{Brandner2000} for a discussion).
Five companion candidates in the higher order systems are very likely chance projections,
with chance projection probabilities between 2\% and 17\%.
These are HD\,158704C, both companion candidates to HD\,178965 (B and C), and the two faint companion candidates in the
quadruple system HD\,66259 (C and D).

\section{Discussion}
\label{sect:discussion}

\begin{figure}
\centering
\includegraphics[width=0.45\textwidth, angle=0]{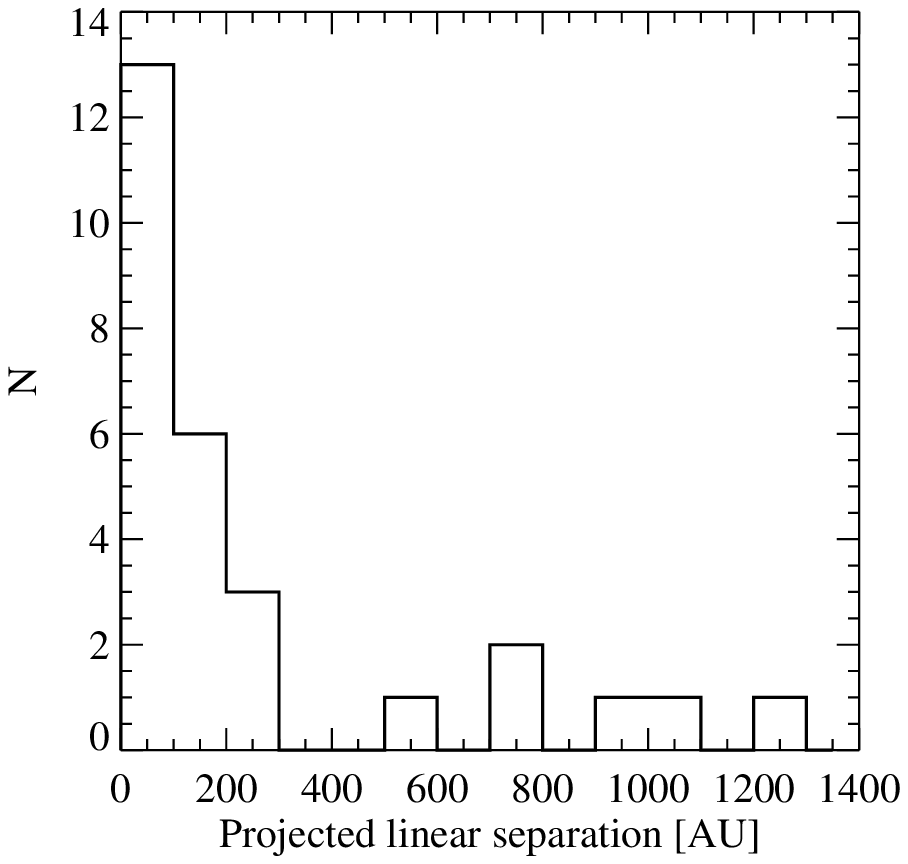}
\caption{
Distribution of the projected separations of the studied systems
with HgMn primaries.
For this figure, we removed all companion candidates with chance projection
probabilities larger than 1\%.
}
\label{fig:histo}
\end{figure}

\begin{table}
\centering
\caption{
Overview of the known multiplicity of the objects studied in this article.
}
\label{tab:multi}
\begin{tabular}{rccc}
\hline
\hline
\multicolumn{1}{c}{HD} &
\multicolumn{1}{c}{SB1} &
\multicolumn{1}{c}{SB2} &
\multicolumn{1}{c}{Astrometric or Visual} \\
\hline
1909   &    & X  &         \\
7374   &  X &    &1        \\
11753  &  X &    &1        \\
14228  &    &    &   1     \\
19400  &    &    &   1     \\
21933  &  x &    &         {$<$1$>$}\\ %
23950  &  x &    &         \\
27295  &  X &    &         \\
27376  &    &  X &      2+{\underline 1}\\ %
28217  &  X &    &   {\underline 1}\\ %
29589  &  x &    &      1  \\
31373  &  x &    &         \\
32964  &    &  X &   1+{\underline 1}\\ %
33647  &    &  X &   {\underline 1}\\ %
33904  &    &    &         {$<$1$>$}\\ %
34364  &    &  X$^a$ &         \\
34880  &    &  X &         {\underline 2}\\ %
35548  &    &  X &   {\underline 1}$^b$\\ %
36881  &  X &    &   {\underline 1}\\ %
37752  &    &    &         \\
38478  &    &    &         \\
42657  &  x &    &   1+{\underline 1}\\ %
49606  &  X &    &   1     \\
51688  &  x &    &   1     \\
53244  &  x &    &         {$<$1$>$}\\ %
53929  &  x &    &         {$<$1$>$}\\ %
59067  &    &    &   3+{\underline 1}\\ %
63975  &    &    &         \\
65949  &  X &    &         \\
65950  &  X &    &         \\
66259  &  X &    &         {$<$1$>$}\\ %
66409  &  X &    &   1     \\
68099  &  x &    &         \\
68826  &    & XX$^d$ &         \\
70235  &    &    &         \\
71066  &    &  x &   3     \\
71833  &    &    &   1     \\
72208  &  X &    &         {$<$1$>$}\\ %
73340  &    &    &      {\underline 1}\\ %
75333  &  X &    &      1+{\underline 1}$^e$\\ %
78316  &  X &    &         {\underline 1}\\ %
90264  &  X &    &         {$<$1$>$}\\ %
101189 &    &    &         {$<$1$>$}\\ %
110073 &  X &    &      {\underline 1}\\ %
120709 &  X &    &   {\underline 1}\\ %
124740 &    &  X &         \\
129174 &  X &    &    1+{\underline 1}\\ %
141556 &  X$^a$ &    &         \\
144661 &  x &    &         \\
144844 &    &  X &   1     \\
158704 &    &  X &   {\underline 1}\\ %
165493 &  X &    &   {\underline 1}\\ %
178065 &    &  X &         \\
216494 &    &  X &   {\underline 1}\\ %
221507 &    &    &         {$<$1$>$}\\ %
224926 &    &    &         \\
\hline
41040  &  X &  X &   {\underline 1}$^c$\\ %
\hline
\end{tabular}  
\begin{flushleft}
Remarks:\\
$^a$There are hints for a third component in these systems.\\
$^b$The visible component is identical with the SB2 system.\\
$^c$The visible component is identical with the SB1 system.\\
$^d$HD\,68826 consists of two SB2 systems.\\
$^e$In fact, we see the two visual components as one.
\end{flushleft}
\end{table}

We have announced the detection of 33 companion candidates in 24 binaries, three triples, and one quadruple system.
The detected companion candidates have K magnitudes between 5\fm95 and 18\fm07 and angular
separations ranging from $<$0\farcs05 to 7\farcs8,
corresponding to linear projected separations of 13.5--1700\,AU.
The companion candidates around HD\,21933, HD\,33904, HD\,53244, HD\,53929, HD\,66259,
HD\,72208, HD\,90264, HD\,101189, and HD\,221507 were detected by us for the first time.
Five companion candidates are very likely to be chance projections.
In Fig.~\ref{fig:histo}, we show the distribution of the projected linear separations for the studied
multiple systems with HgMn primaries.
For half of the systems, the projected linear separations are smaller than 100\,AU.

In our survey, we found in 28 of the 56 studied systems with a HgMn primary
at least one visual companion star, which gives a multiplicity fraction of 50\%.
This is quite high compared with similar surveys of B type stars.
McAlister et al.\ (\cite{McAlister1993}) studied 211 stars of spectral type B with speckle
interferometry at visible wavelengths and obtained a binary fraction of 13.9\%.
Roberts et al.\ (\cite{Roberts2007}) studied 70 B stars in the $I$ band with adaptive optics
and found 16 companions (of which they concluded that four are not physically bound), leading
to a binary fraction $\le$22.9\%.
Duch\^ene et al.\ (\cite{Duchene2001}) surveyed a sample of 60 OB stars in the NGC~6611 cluster with adaptive optics in the $K$ band
and found a binary fraction of 18$\pm$6\%, restricting themselves to a separation range of
0\farcs1 to 1\farcs5, corresponding to 200--3000\,AU, a range where less than half of our
companions are found.
Kouwenhoven et al.\ (\cite{Kouwenhoven2005}) studied the binarity of A and B stars in the
OB association Sco~OB2 with adaptive optics using a $Ks$ filter.
Sixty-five of the 199 stars in their sample have at least one companion, leading to a binary fraction
of 32.7\%.
If one restricts the survey to the 83 B type stars, we find that 23 stars have multiplicity,
giving a multiplicity fraction of 27.7\%.
While our stellar sample is quite heterogeneous in parallax, the parallax of Sco~OB2 from their paper
(they quote a distance of 130\,pc, corresponding to a parallax of 7.7\,mas) is quite similar to the average parallax of our sample
(7.3$\pm$4.4\,mas).
The projected linear separations in their sample (29 to 1600\,AU) are also quite comparable to ours (13.5 to 1700\,AU).
Oudmaijer \& Parr (\cite{OudmaijerParr2010}) observed a sample of 36 B stars and 37 Be stars with NACO on the VLT with
exactly the same camera setting as used in our study.
The only difference is their use of a narrow-band filter, centered on Br$\gamma$, in relation to our $Ks$ filter.
However, their sample is farther away than ours, 4.9$\pm$2.7\,mas for the B stars and 4.4$\pm$2.8\,mas for the Be stars.
They find 21 binaries (10 for the B stars, 11 for the Be stars), leading to a binary fraction of 28.8\%.
Compared to these similar studies, our sample contains
a significantly higher number of stars harboring a companion.

We note that the inspection of SB systems with a late B-type primary in 
the 9$^{\rm th}$ Catalogue of Spectroscopic Binary 
Orbits (Pourbaix et al.\ \cite{Pourbaix2004}) indicates a strong correlation 
between the HgMn peculiarity and membership in a  binary system.
Among the bright well-studied SB systems with late B-type slowly rotating (\vsini{}\,$<$\,70\,km\,s$^{-1}$) primaries
with an apparent magnitude of
up to V$\approx$7  and orbital periods between 3 and 20\,days, apart from HD\,177863, all 21 systems
have a primary with a HgMn peculiarity.
Based on this, it cannot be excluded that 
most late B-type stars formed in binary systems with certain orbital parameters become HgMn stars.  

In Table~\ref{tab:multi}, we present the list of the observed
HgMn stars with notes about their multiplicity.
For each object, we indicate whether it is known to be an SB1 or SB2
and how many astrometric or visual companions are known.
A lower case x indicates that there are hints of an SB system, which
has not yet been confirmed.
Numbers in brackets in the last column indicate objects first
found in this study, while underlined numbers indicate objects
that we were able to confirm.
Of the 56 HgMn stars studied, 32 are confirmed SB systems, 11 are potential SB systems,
and 38 have visual companions.
Only four of the potential SB systems do not have a visual companion.
It is especially intriguing that of the 56 HgMn stars
in the sample studied, only five stars,
HD\,37752, HD\,38478, HD\,63975, HD\,70235, and HD\,224926 
are not known to belong to a binary or multiple system.
This results in a multiplicity rate of 91\%.

In the catalogue of multiple stars by Tokovinin (\cite{Tokovinin1997}),
which compiles data on 728 stellar systems of different spectral types, we
found four additional multiple systems containing HgMn stars.
It is compelling that if
the relative frequency of HgMn stars in multiple systems
is studied, roughly every third system
with a primary in the spectral range between B8 and B9 involves a HgMn star.

The results of our study clearly confirm that HgMn stars are frequently
found in binary and multiple systems.
However, companionship cannot be established based on K photometry alone, and
acquiring data with a near-infrared spectrograph is essential to
establish their true companionship.
Future spectroscopic observations in the near-infrared should be used
to determine the mass of the companions accurately,
and explore the physics in their atmospheres by comparing observed and
synthetic spectra.

We note that our observations contribute not only to the
understanding of the formation mechanism of HgMn stars, but also to the general
understanding of B-type star formation.
An interesting result about the combination of long- and
short-period systems was presented by Tokovinin (\cite{Tokovinin2001}).
He suggested that the fraction of SBs belonging to multiple systems probably depends
on the SB periods.
It is much higher for close binaries with 1 to 10 day periods than
for systems with 10 to 100 day periods.
The statistics of multiple systems is still very poor and much work remains to be done.
The current survey of binarity and multiplicity of HgMn stars will help us to understand the
connection between close binaries and multiplicity, and especially the formation of
close binary systems.
To find out which role membership of HgMn stars in multiple systems plays in
developing their chemical peculiarities, it would be important in the future
to compare the ranges of periods, luminosity ratios, and orbital eccentricities, as well as
the hierarchy of multiples, with the same characteristics in normal late B systems.
In some binary systems with a HgMn primary, the
components definitely rotate subsynchronously (Guthrie \cite{Guthrie1986}).
It is striking that the majority of these systems have more than two components.
Probably the most intriguing and
most fundamental question is whether all late-B close binaries with
subsynchronously rotating companions belong to more complex systems.

\appendix

\section{Notes on individual systems}
\label{sect:individual}

\subsection{Systems unresolved in our study}
\label{sect:unresolved}

{\it HD\,1909:}
This is an SB2 system with a likely period between 5 and 10\,d,
according to Wahlgren et al.\ (\cite{Wahlgren2002}).

{\it HD\,7374:}
This system is an SB1 with a period of 800.9\,d,
according to the 9$^{\rm th}$ Catalogue of Spectroscopic Binary Orbits (Pourbaix et al.\ \cite{Pourbaix2004}).
It is also an astrometric Hipparcos binary according to the CCDM catalogue (Dommanget \& Nys \cite{DommangetNys2002}).

{\it HD\,11753:}
This system is an SB1 with a period of 41.489\,d,
according to the 9$^{\rm th}$ Catalogue of Spectroscopic Binary Orbits (Pourbaix et al.\ \cite{Pourbaix2004}).
It is also an astrometric Hipparcos binary according to the CCDM catalogue (Dommanget \& Nys \cite{DommangetNys2002}).

{\it HD\,14228:}
This star is a fast rotator ($v$\,sin\,$i$=240\,km\,s$^{-1}$; Hutchings et al.\ \cite{Hutchings1979}),
which is atypical for HgMn stars.
It is listed in the ``Catalogue and Bibliography of Mn-Hg Stars'' (Schneider \cite{Schneider1981}).
The Washington Double Star Catalogue
(Mason et al.\ \cite{Mason2001b})
 lists a companion at a distance of 90\arcsec{}.

{\it HD\,19400:}
Dommanget \& Nys (\cite{DommangetNys2002}) mention in the CCDM catalogue a nearby component at a separation of 0\farcs1 and a position angle of 179$^{\circ}$.

{\it HD\,23950:}
This star was marked by Renson \& Manfroid (\cite{RensonManfroid2009})
as an object with variable radial velocity measurements, i.e.\ a potential SB1 system.

{\it HD\,27295:}
This system is an SB1 with a period of 4.4521\,d,
according to the 9$^{\rm th}$ Catalogue of Spectroscopic Binary Orbits (Pourbaix et al.\ \cite{Pourbaix2004}).

{\it HD\,29589:}
Stickland \& Weatherby (\cite{SticklandWeatherby1984}) find hints that HD\,29589 is an SB1 system.
Hubrig et al.\ (\cite{Hubrig2001}) studied this system with ADONIS and found a
companion of K=17.3 at a distance of 10\farcs00 and a position angle of 251.3$^\circ$.
This companion is very likely outside of our effective field-of-view.

{\it HD\,31373:}
This star was marked by Renson \& Manfroid (\cite{RensonManfroid2009})
as an object with variable radial velocity measurements, i.e.\ a potential SB1 system.

{\it HD\,34364:}
This system is an SB2 with a period of 4.1346\,d,
according to the 9$^{\rm th}$ Catalogue of Spectroscopic Binary Orbits (Pourbaix et al.\ \cite{Pourbaix2004}).
Chochol et al.\ (\cite{Chochol1988}) discovered a third body in the system.
The existence of the as yet unseen third star with a mass of at least 0.51\,M$_\odot$ was
inferred from a light-time effect in the observed minima with a period of 25--27\,yr.

{\it HD\,37752:}
There are no references in the literature that indicate multiplicity for this object.

{\it HD\,38478:}
There are no references in the literature that indicate multiplicity for this object.

{\it HD\,49606:}
Stickland \& Weatherby (\cite{SticklandWeatherby1984}) find that HD\,49606 is an SB1 system.
An optical component of magnitude 13.3 at a distance of 27\farcs5 is mentioned by Lindroos (\cite{Lindroos1985}).

{\it HD\,51688:}
This star was marked by Renson \& Manfroid (\cite{RensonManfroid2009})
as an object with variable radial velocity measurements, i.e.\ a potential SB1 system.
Hartkopf et al.\ (\cite{Hartkopf1997}) found a companion on date 1995.3185
at a separation of roughly 0\farcs1 and a position angle of 68.6$^\circ$.
We did not detect this companion.

{\it HD\,63975:}
There are no references in the literature that indicate multiplicity for this object.

{\it HD\,65949:}
Schneider (\cite{Schneider1981}) lists HD\,65949 as an SB1 system.

{\it HD\,65950:}
Abt et al.\ (\cite{Abt1972b}) found variations in the radial velocity of HD\,65950
and classified it as an SB1 system.

{\it HD\,66409:}
Schneider (\cite{Schneider1981}) lists HD\,66409 as an SB1 system.
The Washington Double Star Catalogue
(Mason et al.\ \cite{Mason2001b})
lists a companion at a distance of 12\farcs4.
This companion is very likely outside of our effective field-of-view.

{\it HD\,68099:}
This star was marked by Renson \& Manfroid (\cite{RensonManfroid2009})
as an object with variable radial velocity measurements, i.e.\ a potential SB1 system.

{\it HD\,68826:}
AO\,Vel (HD\,68826) is a quadruple system (Gonz\'alez et al.\ \cite{Gonzalez2006,Gonzalez2010}),
which is composed of two SB2 systems.
The system AB is eclipsing, consists of two BpSi stars, and has an orbital period of 1.584584\,d.
The system CD contains two HgMn stars and has an orbital period of 4.15008\,d.

{\it HD\,70235:}
There are no references in the literature that indicate multiplicity for this object.

{\it HD\,71066:}
This is a quadruple system,
according to the multiple star catalogue of Tokovinin (\cite{Tokovinin1997}).
The brightest component in the system is a potential SB2, with two common proper motion stars at
distances of more than 1\arcmin{} from the primary.
The Washington Double Star Catalogue
(Mason et al.\ \cite{Mason2001b})
lists another component at a separation of
$\sim$35\arcsec{} and a position angle of $\sim$30$^{\circ}$.

{\it HD\,71833:}
Gahm et al.\ (\cite{Gahm1983}) designate a spectral type F2V to
a companion at a separation of 18\farcs9 and V=11.72.

{\it HD\,124740:}
This is an SB2 system,
according to Dolk et al.\ (\cite{Dolk2003}).

{\it HD\,141556:}
This system is an SB2 with a period of 15.2565\,d,
according to the 9$^{\rm th}$ Catalogue of Spectroscopic Binary Orbits (Pourbaix et al.\ \cite{Pourbaix2004}).
HD\,141556 was studied by Hubrig et al.\ (\cite{Hubrig2001}) with ADONIS and
no companion was detected.
The system's $\gamma$-velocity appears to be variable on a timescale
of years, indicating the possible presence of a third body that is invisible
spectroscopically (Dworetsky \cite{Dworetsky1972}).

{\it HD\,144661:}
Significant radial velocity variations were detected by Levato et al.\ (\cite{Levato1987}),
indicating that this object is a potential SB1 system.

{\it HD\,144844:}
Schneider (\cite{Schneider1981}) lists HD\,144844 as an SB2 system.
Our own spectroscopic material also reveals spectral lines of the secondary.
The Washington Double Star Catalogue
(Mason et al.\ \cite{Mason2001b})
lists a companion at a distance of 2\farcs4 and a position angle of 117$^{\circ}$.
We did not detect this companion. 

{\it HD\,224926:}
There are no references in the literature that indicate multiplicity for this object.

\subsection{Systems resolved in our study}
\label{sect:binaries}

{\it HD\,21933:}
This star was marked by Renson \& Manfroid (\cite{RensonManfroid2009})
as an object with variable radial velocity measurements, i.e.\ a potential SB1 system.
We were able to resolve HD\,21933 twice with the newly found companion candidate at a separation of
0\farcs124 and a position angle of 112.2$^{\circ}$ on MJD\,53376.06 
and at a separation of 0\farcs149 and a position angle of 111.9$^{\circ}$ on MJD\,53700.14.

{\it HD\,27376:}
This system is an SB2 with a period of 5.0105\,d,
according to the 9$^{\rm th}$ Catalogue of Spectroscopic Binary Orbits (Pourbaix et al.\ \cite{Pourbaix2004}).
Hubrig et al.\ (\cite{Hubrig2001}) studied this system with ADONIS and found a
companion of K=9.9 at a distance of 5\farcs32 and a position angle of 162.5$^\circ$.
Dommanget \& Nys (\cite{DommangetNys2002}) mention in the CCDM catalogue a nearby component at a separation of 0\farcs1.
The Washington Double Star Catalogue
(Mason et al.\ \cite{Mason2001b})
lists another component at a separation of 49\arcsec{}
and a position angle of 8$^{\circ}$.
We detect the companion found by Hubrig et al.\ (\cite{Hubrig2001}) with K=10.17 at a separation of 5\farcs384 and a position angle of 162.2$^{\circ}$.

{\it HD\,28217:}
This is a triple system,
according to the multiple star catalogue (Tokovinin \cite{Tokovinin1997}).
The main component is an SB1, with a third component 0\farcs4 away from the primary.
The inner orbit has a period of 20.438\,d and the outer orbit of $\sim$170\,yr.
Olevic \& Cvetkovic (\cite{OlevicCvetkovic2005}) communicated an orbital solution for the outer system.
The companion is to be expected on the date of our observations at a separation of 0\farcs104 and a position angle of 359.6$^{\circ}$.
We detect this companion at a separation of 0\farcs119 and a position angle of 27.3$^{\circ}$;
this disagrees with the orbital solution, which needs to be improved.

{\it HD\,32964:}
This is a triple system,
according to the multiple star catalogue (Tokovinin \cite{Tokovinin1997}).
The inner system is an SB2 with a period of 5.5227\,d.
The third component is a faint (V=10.8) star 53\arcsec{} away from the primary.
Hubrig et al.\ (\cite{Hubrig2001}) studied this system with ADONIS and found a
companion of K=9.38 at a distance of 1\farcs613 and a position angle of 232.6$^\circ$.
We detect this companion with K=9.05 at a separation of 1\farcs599 and a position angle of 308.3$^{\circ}$.
Observations in the near future will determine an orbital solution for this fast moving object.

{\it HD\,33647:}
This is a triple system,
according to the Multiple star catalogue (Tokovinin \cite{Tokovinin1997}).
The inner system is an SB2 with a period of 25.365\,d.
The third component in the system is at a separation of 0\farcs109
and has an orbital period of $\sim$120\,yr.
Andrade (\cite{Andrade2004}) communicated an orbital solution for the outer system.
The companion is expected on the date of our observations to be at a separation of 0\farcs145 and a position angle of 346.1$^{\circ}$.
We detect this companion at a separation of 0\farcs147 and a position angle of 345.1$^{\circ}$.

{\it HD\,33904:}
Hubrig et al.\ (\cite{Hubrig2001}) studied HD\,33904 with ADONIS and did not detect a companion.
High spatial resolution X-ray observations of this star revealed that the detected X-rays
do not originate in the B star itself, but rather a previously unresolved companion
(Behar et al.\ \cite{Behar2004}).
Our observations are the first direct detection of a close companion candidate, which we find
at a separation of 0\farcs352 and a position angle of 250.9$^{\circ}$.

{\it HD\,34880:}
Stickland \& Weatherby (\cite{SticklandWeatherby1984}) find that HD\,34880 is an SB2 system.
Dommanget \& Nys (\cite{DommangetNys2002}) mention in the CCDM catalogue a nearby component at a separation of 0\farcs5 and a  position angle of 196$^{\circ}$
and an additional companion at a separation of 4\farcs4 and a position angle of 285$^{\circ}$.
We detect both companions, one at a separation of 0\farcs511 and a position angle of 199.6$^{\circ}$
and the other at a separation of 4\farcs571 and a position angle of 284.9$^{\circ}$.

{\it HD\,35548:}
This is an SB2 system,
according to Dolk et al.\ (\cite{Dolk2003}).
Novakovic (\cite{Novakovic2007}) communicated an orbital solution for this system.
The companion is expected on the date of our observations to be at a separation of 0\farcs292 and a position angle of 174.1$^{\circ}$.
We detect the companion at a separation of 0\farcs300 and a position angle of 174.1$^{\circ}$.

{\it HD\,36881:}
This system is an SB1 with a period of 1857\,d,
according to the 9$^{\rm th}$ Catalogue of Spectroscopic Binary Orbits (Pourbaix et al.\ \cite{Pourbaix2004}).
Dommanget \& Nys (\cite{DommangetNys2002}) mention in the CCDM catalogue a nearby component at a separation of 2\farcs9 and a position angle of 351$^{\circ}$.
We detect this companion at a separation of 2\farcs801 and a position angle of 351.9$^{\circ}$.

{\it HD\,41040:}
Dommanget \& Nys (\cite{DommangetNys2002}) mention in the CCDM catalogue a nearby component at a separation of 0\farcs1 and a position angle of 65$^{\circ}$.
The inner component is an SB2 with a period of 14.57213\,d, the wider component an SB1 system
with a period of 12.98\,yr, which is also visually separated.
Mason et al.\ (\cite{Mason1997}) give an orbital solution for the outer system.
The companion is to be expected on the date of our observations at a separation of 0\farcs055 and a position angle of 59.7$^{\circ}$.
Scarfe et al.\ (\cite{Scarfe2000}) give another orbital solution for the outer system.
The companion is to be expected on the date of our observations at a separation of 0\farcs048 and a position angle of 56.5$^{\circ}$.
We detect the companion to HD\,41040 at a separation of less than 0\farcs050 and a position angle of $\sim$243$^{\circ}$.
Since we measure nearly equal brightness for the two components, we have very likely a 180$^{\circ}$ uncertainty in our measurement.

{\it HD\,42657:}
Aikman (\cite{Aikman1976}) detected small radial velocity variations.
Alzner (\cite{Alzner1998}) gives a separation of 202\farcs8 and a position angle of 5.19$^{\circ}$ for a wide companion.
Dommanget \& Nys (\cite{DommangetNys2002}) mention in the CCDM catalogue a nearby component at a separation of 0\farcs8 and a position angle of 193$^{\circ}$.
We detect this companion at a separation of 0\farcs688 and a position angle of 202.9$^{\circ}$.
Observations in the near future will determine an orbital solution for this relatively fast moving object.

{\it HD\,53244:}
This star was marked by Renson \& Manfroid (\cite{RensonManfroid2009})
as an object with variable radial velocity measurements, i.e.\ a potential SB1 system.
Schneider (\cite{Schneider1981}) also lists HD\,53244 as a potential spectroscopic binary.
We find a new companion candidate to HD\,53244 at a separation of 0\farcs332 and a position angle of 114.8$^{\circ}$.

{\it HD\,53929:}
Radial velocity variations were detected by Zentelis (\cite{Zentelis1983}).
We find a new companion candidate to HD\,53929 at a separation of 3\farcs659 and a position angle of 345.5$^{\circ}$.

{\it HD\,59067:}
McAlister et al.\ (\cite{McAlister1993}) found a companion on date 1988.1731 at a separation of 0\farcs799
and a position angle of 169.0$^{\circ}$.
The Washington Double Star Catalogue
(Mason et al.\ \cite{Mason2001b})
lists three more companions between 20\arcsec{} and 32\arcsec{}.
We find the closest companion at a separation of 0\farcs811 and a position angle of 170.4$^{\circ}$.

{\it HD\,66259:}
Abt et al.\ (\cite{Abt1972b}) found variations in the radial velocity of HD\,66259
and classified it as an SB1 system.
We find a new companion candidate to HD\,66259 at a separation of less than 0\farcs050 and a position angle of $\sim$2.5$^{\circ}$.
The other two sources at a separation of $\sim$5\arcsec{} are very likely to be a chance projection, with
a chance projection probability slightly higher than 5\%.

{\it HD\,72208:}
This system is an SB1 with a period of 22.0116\,d,
according to the 9$^{\rm th}$ Catalogue of Spectroscopic Binary Orbits (Pourbaix et al.\ \cite{Pourbaix2004}).
We find a new companion candidate to HD\,72208 at a separation of 0\farcs671 and a position angle of 332.1$^{\circ}$.

{\it HD\,73340:}
Hubrig et al.\ (\cite{Hubrig2001}) studied this system with ADONIS and found a
companion of K=8.65 at a distance of 0\farcs604 and a position angle of 221.2$^\circ$.
We detect this companion with K=8.54 at a separation of 0\farcs566 and a position angle of 219.7$^{\circ}$.

{\it HD\,75333:}
This target is an SB1 according to the study of Stickland \& Weatherby (\cite{SticklandWeatherby1984}).
Hubrig et al.\ (\cite{Hubrig2001}) studied this system with ADONIS and found a
companion of K=9.45 at a distance of 1\farcs340 and a position angle of 165.8$^\circ$.
In a follow-up campaign at the Keck observatory, Hubrig et al.\ (\cite{Hubrig2005})
resolved the secondary component into another binary.
They found a separation of 1\farcs{}351 for the wide pair A-B
and a separation of about 65$\pm$5\,milliarcseconds for Ba-Bb.
The contrast between A-Ba is about 4.45\,mag and between A-Bb about 5\,mag.
We could not separate the close binary, Ba-Bb, which we detect as one companion with K=9.26 
at a separation of 1\farcs316 and a position angle of 167.5$^{\circ}$.

{\it HD\,78316:}
This system is an SB1 with a period of 6.3933\,d,
according to the 9$^{\rm th}$ Catalogue of Spectroscopic Binary Orbits (Pourbaix et al.\ \cite{Pourbaix2004}).
Mason et al.\ (\cite{Mason2001a}) found a companion on Besselian year 1999.1606 at a separation of 0\farcs286 and a position angle of 108$^{\circ}$.
This star was also resolved by Roberts et al.\  (\cite{Roberts2005}) on Besselian year 2003.0077 with a separation of 0\farcs30
and a position angle of 104${^\circ}$.
We find this companion at a separation of 0\farcs269 and a position angle of 109.7$^{\circ}$.

{\it HD\,90264:}
This is an SB1 system,
according to Dolk et al.\ (\cite{Dolk2003}).
We find a new companion candidate to HD\,90264 at a separation of 2\farcs219 and a position angle of 350.2$^{\circ}$.

{\it HD\,101189:}
There are no references in the literature that indicate multiplicity for this object.
We find a new companion candidate to HD\,101189 at a separation of 0\farcs337 and a position angle of 104.1$^{\circ}$.

{\it HD\,110073:}
This object is an SB1 system,
according to Schneider (\cite{Schneider1981}).
Hubrig et al.\ (\cite{Hubrig2001}) studied this system with ADONIS and found a
companion of K=7.93 at a distance of 1\farcs192 and a position angle of 75.0$^\circ$.
We detect this companion with K=7.96 at a separation of 1\farcs202 and a position angle of 73.9$^{\circ}$.

{\it HD\,120709:}
This is a triple system,
according to the Multiple star catalogue (Tokovinin \cite{Tokovinin1997}).
The inner system is an SB1 with a period of 17.428\,d.
There is a common proper motion object at a distance of 7\farcs851 and a position angle of 106$^\circ$.
We detect this companion at a separation of 7\farcs830 and a position angle of 105.5$^{\circ}$.

{\it HD\,129174:}
Stickland \& Weatherby (\cite{SticklandWeatherby1984}) find that HD\,129174 is an SB1 system.
Dommanget \& Nys (\cite{DommangetNys2002}) mention in the CCDM catalogue a nearby component at a separation of 5\farcs6 and a position angle of 108$^{\circ}$.
The Washington Double Star Catalogue
(Mason et al.\ \cite{Mason2001b})
lists another companion at a separation of 126\arcsec{}.
We detect the close companion at a separation of 5\farcs537 and a position angle of 110.5$^{\circ}$.

{\it HD\,158704:}
This is an SB2 system,
according to Dolk et al.\ (\cite{Dolk2003}).
Hartkopf et al.\ (\cite{Hartkopf1996}) found a companion to HD\,158704
at a separation of 0\farcs352 and a position angle of 11.3$^\circ$ on date 1992.4550.
We find this companion to HD\,158704 at a separation of  0\farcs434 and a position angle of 192.9$^{\circ}$.
The difference in position angle with Hartkopf et al.\ might be due to a 180$^{\circ}$ uncertainty in their measurements.
Another object at a separation of 1\farcs637 and a position angle of 123.6$^{\circ}$ is very likely a
chance projection, with a chance projection probability of $\sim$2\%.

{\it HD\,165493:}
This is an SB1 system,
according to Dolk et al.\ (\cite{Dolk2003}).
Lindroos (\cite{Lindroos1983}) gives a separation of 3\farcs9 for a companion in the system.
We detect this companion at a separation of 4\farcs041 and a position angle of 257.4$^{\circ}$.

{\it HD\,178065:}
This target is an SB1 system with an orbital period of 6.87\,d according to Guthrie (\cite{Guthrie1984}).
Dolk et al.\ (\cite{Dolk2003}) identified HD\,178065 as an SB2 system.
Mason et al.\ (\cite{Mason1999}) list HD\,178065 as an unresolved Hipparcos problem star.
The two sources found by us close to HD\,178065 at 6\farcs4 and 3\farcs3 are very likely to be a chance projection,
with a chance projection probability of 17\%.

{\it HD\,216494:}
This is a triple system,
according to the Multiple star catalogue (Tokovinin \cite{Tokovinin1997}).
The inner system is an SB2 with a period of 3.4298\,d.
An occulting, visual component on an 18\,yr orbit was found at a separation of 0\farcs078.
Mason (\cite{Mason1997}) communicated an orbital solution for the outer system.
The companion is to be expected on the date of our observations at a separation of 0\farcs080 and a position angle of 288.9$^{\circ}$.
We detect this companion at a separation of 0\farcs069 and a position angle of 285.9$^{\circ}$.

{\it HD\,221507:}
Hubrig et al.\ (\cite{Hubrig2001}) studied HD\,221507 with ADONIS and did not
detect a companion.
We find a new companion candidate to HD\,221507 at a separation of 0\farcs641 and a position angle of 240.2$^{\circ}$.

\begin{acknowledgements}

We would like to thank Christian Hummel for his help calculating the binary
positions from published orbital elements, and the referee, Douglas Gies, for
his valuable comments, which improved the quality of this paper.
Part of this work was supported by the ESO Director General
Discretionary Fund.
This publication made use of data products from the Two Micron All Sky Survey,
the DENIS database,
and of the SIMBAD database, operated at CDS, Strasbourg, France.

\end{acknowledgements}


\begin{thebibliography}{}

\bibitem[1972]{Abt1972a}
Abt, H.~A., Chaffee, F.~H., \& Suffolk, G.\ 1972,
ApJ, 175, 779

\bibitem[1972]{Abt1972b}
Abt, H.~A., \& Levy, S.~G.\ 1972,
\apj, 172, 355

\bibitem[2002]{Adelman2002}
Adelman, S.~J., Gulliver, A.~F., Kochukhov, O.~P., \& Ryabchikova, T.~A.\ 2002,
\apj, 575, 449

\bibitem[1976]{Aikman1976}
Aikman, G.~C.~L.\ 1976,
Publ.\ of the Dominion Astrophys.\ Obs.\ Victoria, 14, 379

\bibitem[1998]{Alzner1998}
Alzner, A.\ 1998,
A\&AS, 132, 237

\bibitem[2004]{Andrade2004}
Andrade, M.\ 2004,
IAU Commission on Double Stars, 154, 1

\bibitem[2004]{Behar2004}
Behar, E., Leutenegger, M., Doron, R., et al.\ 2004,
ApJL, 612, L65

\bibitem[1996]{Berghoefer1996}
Bergh\"ofer, T.~W., Schmitt, J.~H.~M.~M., \& Cassinelli, J.~P.\ 1996,
A\&AS, 118, 481

\bibitem[2000]{Brandner2000}
Brandner, W., Zinnecker, H., Alcal\'a, J.~M., et al.\ 2000,
AJ, 120, 950

\bibitem[2010]{Briquet2010}
Briquet, M., Korhonen, H., Gonz{\'a}lez, J.~F., et al.\ 2010,
A\&A, 511, A71

\bibitem[1997]{Castelli1997} 
Castelli, F., Parthasarathy, M., \& Hack, M.\ 1997,
A\&A, 321, 254

\bibitem[1988]{Chochol1988} 
Chochol, D., Juza, K., Zverko, J., et al.\ 1988,
Bull.\ of the Astr.\ Inst.\ of Czechoslovakia, 39, 69

\bibitem[1992]{Cole1992}
Cole, W.~A., Fekel, F.~C., Hartkopf, W.~I., et~al.\ 1992,
AJ, 103, 1357

\bibitem[2006]{Correia2006}
Correia, S., Zinnecker, H., Ratzka, Th., \& Sterzik, M.~F.\ 2006,
A\&A, 459, 909

\bibitem[2003]{Dolk2003}
Dolk, L., Wahlgren, G.~M., \& Hubrig, S.\ 2003,
\aap, 402, 299

\bibitem[2002]{DommangetNys2002}
Dommanget, J., \&  Nys, O.\ 2002,
CCDM (Catalog of Components of Double \& Multiple stars)

\bibitem[2001]{Duchene2001}
Duch{\^e}ne, G., Simon, T., Eisl{\"o}ffel, J., \& Bouvier, J.\ 2001,
A\&A, 379, 147

\bibitem[1972]{Dworetsky1972}
Dworetsky, M.~M.\ 1972,
PASP, 84, 254

\bibitem[1983]{Gahm1983}
Gahm, G.~F., Ahlin, P., \& Lindroos, K.~P.\ 1983,
\aaps, 51, 143

\bibitem[2006]{Gonzalez2006}
Gonz\'alez, J.~F., Hubrig, S., Nesvacil, N., \& North, P.\ 2006,
A\&A, 449, 327

\bibitem[2010]{Gonzalez2010}
Gonz{\'a}lez, J.~F., Hubrig, S., \& Castelli, F.\ 2010,
MNRAS, 402, 2539

\bibitem[1984]{Guthrie1984}
Guthrie, B.~N.~G.\ 1984,
MNRAS, 206, 85

\bibitem[1986]{Guthrie1986}
Guthrie, B.~N.~G.\ 1986,
MNRAS, 220, 559

\bibitem[1996]{Hartkopf1996}
Hartkopf, W.~I., Mason, B.~D., McAlister, H.~A., et al.\ 1996,
\aj, 111, 936

\bibitem[1997]{Hartkopf1997}
Hartkopf, W.~I., McAlister, H.~A., Mason, B.~D., et al.\ 1997,
\aj, 114, 1639

\bibitem[1995]{HubrigMathys1995}
Hubrig, S., \& Mathys, G.\ 1995,
Comm.\ on Astroph., 18, 167

\bibitem[1996]{HubrigMathys1996}
Hubrig, S., \& Mathys, G.\ 1996, 
A\&A, 120, 457

\bibitem[1998]{HubrigBerghoefer1998}
Hubrig, S., \& Bergh\"ofer, T.~W.\ 1998,
in: ``The Hot Universe'',
IAU Symposium 188,
eds.\ K.~Koyama, S.~Kitamoto, \& M.~Itoh, p.\ 217

\bibitem[2001]{Hubrig2001}
Hubrig, S., Le Mignant, D., North, P., \& Krautter, J.\ 2001,
A\&A, 372, 152

\bibitem[2005]{Hubrig2005}
Hubrig, S., Sch{\"o}ller, M., Le Mignant, D., et al.\ 2005,
in: ``High Resolution Infrared Spectroscopy in Astronomy'',
eds.\ H.~U.\ K{\"a}ufl, R.\ Siebenmorgen, \& A.\ Moorwood, p.\ 499

\bibitem[2006]{Hubrig2006}
Hubrig, S., Gonz{\'a}lez, J.~F., Savanov, I., et al.\ 2006,
\mnras, 371, 1953

\bibitem[1979]{Hutchings1979}
Hutchings, J.~B., Nemec, J.~M., \& Cassidy, J.\ 1979,
PASP, 91, 313

\bibitem[1991]{Isobe1991}
Isobe, S.\ 1991,
Proc.\ ASA, 9, 270

\bibitem[2007]{Kochukhov2007}
Kochukhov, O., Adelman, S.~J., Gulliver, A.~F., \& Piskunov, N.\ 2007,
Nature Physics, 3, 526

\bibitem[2005]{Kouwenhoven2005}
Kouwenhoven, M.~B.~N., Brown, A.~G.~A., Zinnecker, H., et al.\ 2005,
\aap, 430, 137

\bibitem[2003]{Lenzen2003}
Lenzen, R., Hartung, M., Brandner, W., et al.\ 2003,
SPIE Conf.\ Ser.\ 4841, p.\ 944

\bibitem[1987]{Levato1987}
Levato, H., Malaroda, S., Morrell, N., \& Solivella, G.\ 1987,
ApJS, 64, 487

\bibitem[1983]{Lindroos1983}
Lindroos, K.~P.\ 1983,
A\&AS, 51, 161

\bibitem[1985]{Lindroos1985}
Lindroos, K.~P.\ 1985,
A\&AS, 60, 183

\bibitem[2003]{Masciadri2003}
Masciadri, E., Brandner, W., Bouy, H., et al.\ 2003,
A\&A, 411, 157

\bibitem[1997]{Mason1997}
Mason, B.~D.\ 1997,
\aj, 114, 808

\bibitem[1999]{Mason1999}
Mason, B.~D., Martin, C., Hartkopf, W.~I., et al.\ 1999,
AJ, 117, 1890

\bibitem[2001a]{Mason2001a}
Mason, B.~D., Hartkopf, W.~I., Holdenried, E.~R., \& Rafferty, T.~J.\ 2001a,
AJ, 121, 3224

\bibitem[2001b]{Mason2001b}
Mason, B.~D., Wycoff, G.~L., Hartkopf, W.~I., et al.\ 2001b,
AJ, 122, 3466

\bibitem[1993]{McAlister1993}
McAlister, H.~A., Mason, B.~D., Hartkopf, W.~I., \& Shara, M.~M.\ 1993,
AJ, 106, 1639

\bibitem[1994]{NordstromJohansen1994}
Nordstrom, B., \& Johansen, K.~T.\ 1994,
\aap, 282, 787

\bibitem[2007]{Novakovic2007}
Novakovic, B.\ 2007,
Chin.\ J.\ of Astronomy and Astrophysics, 7, 415

\bibitem[2005]{OlevicCvetkovic2005}
Olevic, D., \& Cvetkovic, Z.\ 2005,
IAU Commission on Double Stars, 155, 1

\bibitem[2010]{OudmaijerParr2010}
Oudmaijer, R.~D., \& Parr, A.~M.\ 2010,
MNRAS, {\sl in print}, also: arXiv:1003.0618

\bibitem[2004]{Pourbaix2004}
Pourbaix, D., Tokovinin, A.~A., Batten, A.~H., et al.\ 2004,
\aap, 424, 727

\bibitem[2009]{RensonManfroid2009}
Renson, P., \& Manfroid, J.\ 2009,
\aap, 498, 961

\bibitem[2005]{Roberts2005}
Roberts, L.~C., Jr., Turner, N.~H., Bradford, L.~W., et al.\ 2005,
AJ, 130, 2262

\bibitem[2007]{Roberts2007}
Roberts, Jr., L.~C., Turner, N.~H., \& ten Brummelaar, T.~A.\ 2007,
AJ, 133, 545

\bibitem[2003]{Rousset2003}
Rousset, G., Lacombe, F., Puget, P., et al.\ 2003,
SPIE Conf.\ Ser.\ 4839, p.\ 140

\bibitem[2000]{Scarfe2000}
Scarfe, C.~D., Barlow, D.~J., \& Fekel, F.~C.\ 2000,
AJ, 119, 2415

\bibitem[1981]{Schneider1981}
Schneider, H.\ 1981,
\aaps, 44, 137

\bibitem[1984]{SticklandWeatherby1984}
Stickland, D.~J., \& Weatherby, J.\ 1984,
\aaps, 57, 55

\bibitem[1997]{Tokovinin1997}
Tokovinin, A.\ 1997,
A\&AS, 124, 75

\bibitem[2001]{Tokovinin2001}
Tokovinin, A.\ 2001,
in: ``The Formation of Binary Stars'',
IAU Symposium 200,
eds.\ H.~Zinnecker \& R.~Mathieu, p.\ 84

\bibitem[2002]{Wahlgren2002}
Wahlgren, G.~M., Hubrig, S., \& Dolk, L.\ 2002,
Information Bulletin on Variable Stars, 5290, 1

\bibitem[1974]{WolffWolff1974}
Wolff, S.~C., \& Wolff, R.~J.\ 1974,
ApJ, 194, 65

\bibitem[1983]{Zentelis1983}
Zentelis, N.\ 1983,
A\&AS, 53, 445

\end{thebibliography}
\end{document}